\documentclass[]{elsarticle}
\usepackage{amssymb}
\usepackage{amsmath}
\usepackage{graphicx}
\usepackage{subfigure}
\usepackage{epstopdf}
\usepackage{color}
\usepackage{multicol}

\begin{document}

\begin{frontmatter}

\title{Stability of solitons in time-modulated two-dimensional lattices}
\author{Nir Dror$^1$ and Boris A. Malomed$^{1,2}$}
\address{$^1$Department of Physical Electronics, School of Electrical
Engineering, Faculty of Engineering,Tel Aviv University, Tel Aviv
69978, Israel\\
$^2$ITMO University, St. Petersburg 197101, Russia}

\begin{abstract}
We develop stability analysis for matter-wave solitons in a two-dimensional
(2D) Bose-Einstein condensate loaded in an optical lattice (OL), to which
periodic time modulation is applied, in different forms. The stability is
studied by dint of the variational approximation and systematic simulations.
For solitons in the semi-infinite gap, well-defined stability patterns are
produced under the action of the attractive nonlinearity, clearly exhibiting
the presence of resonance frequencies. The analysis is reported for several
time-modulation formats, including the case of in-phase modulations of both
quasi-1D sublattices, which build the 2D square-shaped OL, and setups with
asynchronous modulation of the sublattices. In particular, when the
modulations of two sublattices are phase-shifted by $\delta =\pi /2$, the
stability map is not improved, as the originally well-structured stability
pattern becomes fuzzy and the stability at high modulation frequencies is considerably reduced.
Mixed results are obtained for anti-phase modulations of the sublattices ($\delta =\pi$), where
extended stability regions are found for low modulation frequencies, but for high frequencies
the stability is weakened. The
analysis is also performed in the case of the repulsive nonlinearity,
for solitons in the first finite bandgap. It is concluded that, even though
stability regions may be found, distinct stability boundaries for the gap
solitons cannot be identified clearly. Finally, the stability is also
explored for vortex solitons of both the ``square-shaped"
and ``rhombic" types (i.e., off- and on-site-centered ones).
\end{abstract}

\end{frontmatter}

\section{Introduction}

The study of matter-wave solitons in Bose-Einstein Condensates (BECs), and
in particular in multidimensional settings, has attracted a great deal of
attention in the two recent decades. In the one-dimensional (1D) setting,
solitons are known to be stable in condensates with attractive interactions
between atoms. In experiments, effectively 1D matter-wave solitons were
observed when using cigar-shaped (elongated) harmonic traps in condensates
of $^{7}$Li \cite{Strecker1,Khaykovich,Strecker2} and $^{85}$Rb \cite%
{Cornish} atoms. In higher dimensions, however, the attractive force between
the atoms cannot balance the kinetic pressure, which causes instability of
the condensate against the critical collapse in 2D, and supercritical
collapse in 3D \cite{Berge,Fibich}.

Various approaches to controlling the dynamics of matter waves in BECs and,
in the context of the present work, stabilizing multidimensional solitons,
were proposed. A ubiquitous stabilization technique may be provided by
optical lattices (OLs), which are induced, in experiments, as interference
patterns, through coherent laser beams illuminating the condensate in
opposite directions \cite{Morsch}. The spatially periodic distribution of
the light intensity in the OL gives rise to an effective spatially periodic
potential applied to the boson gas. Intensive theoretical analysis has
predicted that, in the case of attractive interactions between atoms, 2D and
3D OLs \cite{Baizakov1,Efremidis,Musslimani,Musslimani1,Baizakov3} may stabilize
solitons of the same dimension against the collapse. In addition,
low-dimensional OLs, which are expressed as periodic potentials whose
dimension is smaller by 1 from that of the embedding space, can also support
stable 2D and 3D solitons \cite{Baizakov3,Mihalache}.

In BEC with repulsive interactions between atoms, which is the generic case
\cite{Pethik}, bright solitons do not exist in the free space, but they can
be supported by OLs, in the form of gap solitons (GSs) inside finite
spectral bandgaps induced by the OL potential. The concept behind the
formation of the GS is that the sign of the effective mass of collective
excitations may be inverted under the action of the lattice potential, which
thus balances the repulsive nonlinearity. Fundamental GSs were studied in
both 1D \cite{Abdullaev} and multidimensional \cite%
{Baizakov0,Louis,Efremidis1,Ostrovskaya0,Sakaguchi,MultidimensionalGS2,Gubeskys2,Shi}
geometries. In the experiment, effectively 1D GSs, composed of a few
hundreds of atoms, were created in $^{87}$Rb condensate, loaded into the OL
potential \cite{Eiermann}.

Besides the fundamental solitons and GSs, other types of multidimensional
modes were also studied. In particular, many works dealt with vortex
solitons i.e., multi-peak ring-shaped structures with embedded global
vorticity imprinted onto the ring complex, in the semi-infinite gap, under
the attractive nonlinearity \cite%
{Baizakov1,Musslimani,Neshev,Fleischer,Alexander,Gubeskys1,Richter,Mayteevarunyoo1,Wang}%
, as well as gap vortex solitons under the repulsive nonlinearity \cite%
{MultidimensionalGS2,Sakaguchi,Gubeskys1,Richter,Mayteevarunyoo1,Wang,Ostrovskaya}%
. The vortex solitons with the most basic structure are composed of four
density peaks and may be classified into two categories: square-shaped,
alias off-site-centered vortices, which are built as dense patterns with the
central point positioned at a local maximum of the OL potential \cite%
{Musslimani,Neshev,Fleischer,Alexander,MultidimensionalGS2,Mayteevarunyoo1,Wang,Ostrovskaya}%
, and rhombic configurations, alias \textquotedblleft diamonds" or
on-site-centered vortices, which feature a vacant lattice site at the center
\cite{Fleischer,Alexander,Gubeskys1,Richter,Mayteevarunyoo1,Wang,Ostrovskaya}%
.

Another effective means for controlling the dynamics of the BEC may be
provided by subjecting various parameters, which affect properties of the
condensate, to periodic time-modulation (these tools belong to the general
class of \textit{management techniques} \cite{SolitonManagement}). One such
technique involves periodic time modulation of the strength of the potential
trap, which confines either repulsive \cite%
{GarciaRipoll1,GarciaRipoll2,Abdullaev2,Abdullaev3} or attractive \cite%
{Abdullaev4,Baizakov2} condensates. Varying the amplitude and frequency of
the temporal modulation may reveal possibilities of creating parametric
resonances in the BEC \cite{Baizakov2}. Also belonging to this group of
techniques is the periodic modulation of the nonlinearity strength, achieved
via temporal modification of the scattering length of atomic collisions.
This effect may be provided through the \emph{Feshbach-resonance management}
(FRM), induced by a low-frequency ac magnetic field applied to the
condensate. It was predicted that the FRM is capable of stabilizing 2D
solitons \cite{FRM2DSolitons}, as well as 2D vortices (but not 3D solitons),
in free space. The stability of 3D solitons and their bound complexes was
also examined, under the combined action of the FRM and a quasi-1D OL
potential \cite{Staliunas,Trippenbach}. In the 1D setting, for the
condensate trapped in the static parabolic potential, the FRM scheme gives
rise to dynamical states such as breathers and stable two-soliton bound
structures \cite{Kevrekidis}. The dynamics of 3D solitons was also studied
in BEC where both the cubic and the quintic nonlinear terms are periodically
modulated in time (i.e., both two- and three-body interactions are
considered) \cite{Sabari1}.

A different management approach for BEC loaded in the OL potential relies
upon periodic time modulation of the lattice's strength. In the framework of
this \emph{lattice-management} technique, especially interesting are cases
when the OL is necessary for the existence or stability of the solitons. The
stability of both fundamental GSs (in the first and second bandgaps) and
their bound states, in the framework of the 1D Gross-Pitaevskii equation
(GPE), with a repulsive cubic term, was explored in Refs. \cite%
{Mayteevarunyoo1D1,Mayteevarunyoo1D2}. In 2D models, similar analyses were
reported, in the case of the attractive nonlinearity, for quasi-1D \cite%
{MayteevarunyooQuasi1D} and full 2D OL \cite{Burlak} potentials. The latter
work did not examine the effect of the modulation frequency on the
long-lived stability of the solitons, an issue which is considered in the
present work. Furthermore, while the analysis performed in previous works
was limited to the management format with synchronous modulation of both 1D
sublattices building a square-shaped 2D lattice, we extend the analysis to
other formats, with asynchronous modulations applied to the sublattices. The
stability investigation for vortex solitons, of both the square and rhombus
types, and 2D GSs in the first finite bandgap (under the action of the
repulsive nonlinearity, in the latter case), are reported too.

The rest of the paper is structured as follows. The model is formulated is
Sec. ~\ref{sec:model}. In Sec.~\ref{sec:VA} we formulate the variational
approximation (VA) to develop an analytical description of the dynamics of
fundamental solitons in the 2D\ GPE, which incudes the general expression
for the time-modulated OL, see Eqs. (\ref{2DGPE}) and (\ref%
{General2DExternalPotential}) below. The stability results, for the models
with the attractive and repulsive nonlinearities, under the action of the
isotropic time-modulated OL, are reported in Sec.~\ref%
{sec:SymmetricModulation}, using both the VA and direct numerical
computations. In Sec.~\ref{sec:NonsynchronousModulation}, similar stability
analysis is reported for the modulation formats which are asynchronous with
respect to the 1D sublattices. The stability of square- and rhombic-shaped
vortices is considered in Sec.~\ref{sec:VorticesStyability}. The paper is
concluded in Sec.~\ref{sec:Conclusions}.

\label{sec:Introduction}

\section{The model}

\label{sec:model} We start with the 3D GPE, which governs the dynamics of
atomic BEC in the mean-field approximation:

\begin{equation}
i\hbar \frac{\partial \Psi }{\partial T}=\left[ -\frac{\hbar }{2m}\nabla
^{2}+W(X,Y,Z)+\frac{4\pi \hbar ^{2}a_{s}N}{m}|\Psi |^{2}\right] \Psi .
\label{3DGPE}
\end{equation}%
Here $\Psi (\mathbf{R},T)$ is the wave function of the condensate at
position $\mathbf{R}$ and time $T$, subject to the normalization condition $%
\int |\Psi (\mathbf{R},T)|^{2}d\mathbf{R}=1$, $N$ is the total number of
atoms in the condensate, $m$ the atomic mass and $a_{s}$ the \textit{s}-wave
scattering length, with $a_{s}<0$ and $a_{s}>0$ referring to self-attractive
and repulsive condensates, respectively. In the context of the present work,
the external potential, $W(X,Y,Z)$ (for the time being, it is taken in the
time-independent form), includes the confining harmonic-oscillator term,
acting in the $Z$ direction, and the 2D OL with half-depth $W_{0}$ and
period $d$, in the $\left( X.Y\right) $ plane:

\begin{equation}
W(X,Y,Z)=\frac{1}{2}m\omega _{z}^{2}Z^{2}-W_{0}\left[ \cos \left( \frac{2\pi
X}{d}\right) +\cos \left( \frac{2\pi Y}{d}\right) \right]
\label{Confined2DPeriodicPotential}
\end{equation}

Reducing the 3D system to a 2D equation is performed by means of the usual
method, substituting $\Psi (X,Y,Z,T)=\Phi(X,Y,T)\exp \left( -\frac{i}{2}\omega
_{z}T-\frac{Z^{2}}{2a_{z}^{2}}\right) $, where $a_{z}$ is the
transverse-confinement length, $a_{z}=\sqrt{\hbar /m\omega _{z}}$. Further,
the original variables are rescaled as follows: $(X,Y)=(d/\pi )(x,y)$, $%
T=(md^{2}/\pi ^{2}\hbar )t$, $W_{0}=V_{0}(\pi ^{2}\hbar ^{2}/md^{2})$, $\Phi
=d^{-1}\left(\sqrt{\pi /2\sqrt{2}|a_{s}|}\right)U$. The resulting rescaled form of the
2D equation is

\begin{equation}
i\frac{\partial U}{\partial t}=-\frac{1}{2}\left( \frac{\partial ^{2}U}{%
\partial x^{2}}+\frac{\partial ^{2}U}{\partial y^{2}}\right) +\sigma
|U|^{2}U+V(x,y)U,  \label{2DGPE}
\end{equation}%
where $\sigma =+1$ and $-1$ for the repulsive and attractive nonlinearities,
respectively, and the rescaled external potential is

\begin{equation}
V(x,y)=-V_{0}[\cos (2x)+\cos (2y)]  \label{Basic2DExternalPotential}
\end{equation}

As said above, the subject of the present work is periodic time modulation
of the OL. For this purpose, static OL potential (\ref%
{Basic2DExternalPotential}) is replaced by

\begin{equation}
V(x,y;t)=-V_{0}\left\{ \left[ \kappa +\frac{\varepsilon }{2}\sin \left(
\omega t \right) \right] \cos (2x)+\left[ \kappa +\frac{\varepsilon }{2}\sin
\left( \omega t+\delta\right) \right] \cos (2y)\right\} .
\label{General2DExternalPotential}
\end{equation}

In particular, $\delta =\pi /2$ corresponds to the combination of the usual
square-shaped static lattice potential $\sim \kappa $ and rotating one $\sim
\varepsilon $ in Eq. (\ref{General2DExternalPotential}). We first examine
the case of the synchronously modulated square-shaped lattice, with $\delta =0$. We
then apply obvious rescaling to set $\kappa =1$ in Eq. (\ref%
{General2DExternalPotential}), unless the case of $\kappa =0$ is considered.
Note that, at
\begin{equation}
\varepsilon =2\kappa \equiv 2,  \label{2}
\end{equation}%
the potential periodically switches between the strongest OL with amplitude $%
2V_{0}$ (at $\omega t=\pi /2+2\pi n$) and the free-space configuration with
no OL, at $\omega t=3\pi /2+2\pi n$, where $n$ is an arbitrary integer.

Next, we will examine scenarios with non-zero phase shifts, $\delta \neq 0$,
between the temporal modulations acting on the 1D sublattices in Eq. (\ref%
{General2DExternalPotential}). In that case, even if condition (\ref{2}) is
imposed, some form of the OL potential is present at all times, hence
stability of 2D solitons, which is supported by the lattice potential, may
be expected to be stronger. In this work, two nonzero phase shifts are
considered, $\delta =\pi /2$ and $\pi $. In the latter case (the anti-phase
modulation of the sublattices), under condition (\ref{2}), the OL potential
periodically alternates between quasi-1D OLs acting along axes $x$ and $y$.

Stability regions for 2D solitons supported by the time-modulated OLs are
identified below by means of numerical methods in all the above-mentioned
cases, and the results are compared with those predicted by the
variational approximation (VA). We pay particular attention to resonances
that may be found under the action of different time-modulation patterns. In
the system considered in the present work, resonant frequencies appear when
the time-modulation frequency coincides with a multiple of the fundamental
frequency of collective oscillations in the trapped BEC. A fairly good
assessment of the resonant frequencies may be achieved by plotting stability
maps in the plane of the modulation parameters, $(\omega ,\varepsilon )$,
and looking for the base points from which instability tongues originate, if
any. This analysis is performed for several settings considered in the
present work, using both systematic simulations of Eq. (\ref{2DGPE}) and VA
method.

In addition to that, we consider the special case when the OL potential does
not contain a static component, i.e., $\kappa =0$ in Eq. (\ref%
{General2DExternalPotential}). Actually, in this case all the solitons are
found to be \emph{unstable}, irrespective of the value of phase shift $%
\delta $.

Finally, the stability investigation is performed for solutions different
from the 2D fundamental solitons, \textit{viz}., families of four-peak
square-shaped and rhombic vortices, in the case of the isotropic temporal
modulation.

Results for the stability, presented below, were collected by means of
systematic direct simulations, using the standard pseudospectral split-step
Fourier method. 2D stationary soliton solutions in the static OL, in the
form of
\begin{equation}
u\left( x,y,t\right) =\exp \left( -i\mu t\right) U\left( x,y\right) ,
\label{mu}
\end{equation}%
where $\mu $ is a real chemical potential, were used as initial conditions
for the dynamical simulations. These stationary solutions were obtained by
means of the modified squared-operator method \cite{Lakoba}.

All the stability diagrams presented in this work exhibit the results in a
confined modulation-frequency region, $0<\omega <8$. This range allows
sufficiently clear observation of rapid changes occurring at low modulation
frequencies and, on the other hand, makes it possible to capture trends at
higher frequencies. If needed, additional results, obtained for $\omega >8$,
are mentioned too.

\section{The variational approximation}

\label{sec:VA}

The VA can be applied to the model at hand, similar to how it was done in
Ref. \cite{Burlak}. The Lagrangian corresponding to Eq. (\ref{2DGPE}) is $%
L=\int_{-\infty }^{+\infty }\mathcal{L}dxdy$, with density

\begin{gather}
\mathcal{L}=\frac{i}{2}(u^{\ast }u_{t}-uu_{t}^{\ast })-\frac{1}{2}%
(|u_{x}|^{2}+|u_{y}|^{2})-\sigma \frac{1}{2}|u|^{4}  \notag \\
+V_{0}\left\{ \left[ \kappa +\frac{\varepsilon }{2}\sin \left( \omega
t\right) \right] \cos (2x)+\left[ \kappa +\frac{\varepsilon }{2}\sin \left(
\omega t+\delta \right) \right] \cos (2y)\right\} |u|^{2}.
\label{LagrangianDensity}
\end{gather}%
We here chose the commonly used Gaussian ansatz, written as

\begin{equation}
u(x,y,t)=A(t)\exp(i\varphi(t))\prod_{\eta =x,y}\exp \left[ \frac{i}{2}%
b_{\eta }(t)\eta ^{2}-\frac{\eta ^{2}}{2W_{\eta }^{2}(t)}\right] ,
\label{GaussianSolution}
\end{equation}%
with real amplitude $A$, overall phase $\varphi$, widths $W_{\eta }$, chirps
$b_{\eta }$, and norm $N=\pi A^{2}W_{x}W_{y}$. Substituting ansatz (\ref%
{GaussianSolution}) in Eq. (\ref{LagrangianDensity}) and calculating the
integrals results in the following \emph{effective Lagrangian}:

\begin{equation}
\begin{split}
L_{\mathrm{eff}}& =-N\frac{d\varphi }{dt}-\frac{1}{4}N\left( \frac{1}{%
W_{x}^{2}}+\frac{1}{W_{y}^{2}}\right) -\frac{\sigma N^{2}}{4\pi W_{x}W_{y}}
\\
& +V_{0}N\left\{ \left[ \kappa +\frac{\varepsilon }{2}\sin \left( \omega
t\right) \right] e^{-W_{x}^{2}}+\left[ \kappa +\frac{\varepsilon }{2}\sin
\left( \omega t+\delta \right) \right] e^{-W_{y}^{2}}\right\} \\
& -\frac{1}{4}\frac{db_{x}}{dt}NW_{x}^{2}-\frac{1}{4}\frac{db_{y}}{dt}%
NW_{y}^{2}-\frac{1}{4}N(b_{x}^{2}W_{x}^{2}+b_{y}^{2}W_{y}^{2}),
\end{split}%
\end{equation}%
The next step is solving the Euler-Lagrange equations for the variational
parameters, $A$, $W_{\eta }$, $b_{\eta }$, $\varphi $. The equation for
phase $\varphi $ amounts to the conservation of the norm: $dN/dt=0$. The
variational equations for the chirps express $b_{\eta }$ in terms of the
time derivatives of the widths: $b_{\eta }=W_{\eta }^{-1}dW_{\eta }/dt$.
Finally, the equations for the widths produce an eventual system of
dynamical equations:
\begin{equation}
\begin{split}
\frac{d^{2}W_{x}}{dt^{2}}& =\frac{1}{W_{x}^{2}}\left( \frac{1}{W_{x}}+\sigma
\frac{N/2\pi }{W_{y}}\right) -4W_{x}V_{0}\left\{ \left[ \kappa +\frac{%
\varepsilon }{2}\sin \left( \omega t\right) \right] e^{-W_{x}^{2}}\right\} ,
\\
\frac{d^{2}W_{y}}{dt^{2}}& =\frac{1}{W_{y}^{2}}\left( \frac{1}{W_{y}}+\sigma
\frac{N/2\pi }{W_{x}}\right) -4W_{y}V_{0}\left\{ \left[ \kappa +\frac{%
\varepsilon }{2}\sin \left( \omega t+\delta \right) \right]
e^{-W_{y}^{2}}\right\} .
\end{split}
\label{WidthVaEquations}
\end{equation}

In the subsequent sections Eqs. (\ref{WidthVaEquations}) are utilized to
predict stability of the dynamical soliton modes. Simulations of Eq. (\ref%
{WidthVaEquations}) were performed by means of a simple finite-difference
scheme.

\section{The 2D lattice under the isotropic time modulation}

\label{sec:SymmetricModulation}

In this section we consider the 2D square-shaped OL with the synchronized time
modulation applied to both 1D\ sublattices, which corresponds to $\delta =0$
in Eqs. (\ref{General2DExternalPotential}) and, eventually, in Eq. (\ref%
{WidthVaEquations}).
First, we address a particular case, with $\kappa =0$ and $\varepsilon \neq 0$, when the static component is absent in the OL. The detailed analysis based on the VA, as well as systematic direct simulations of the full GPE (\ref{2DGPE}) in a wide range of initial conditions, lead to a conclusion that the model without the static part of the OL potential cannot sustain stable soliton-like solutions, for either sign of $\sigma $ (attractive and repulsive nonlinearities).
This outcome is understandable because, for all values of the phase shift besides $\delta=\pi$, both $x$- and $y$-sublattices that form the trapping potential (\ref{General2DExternalPotential}) switch their signs in a part of the modulation period (half of the period, for the synchronous setting, $\delta=0$), which tends to destroy
the soliton. Our numerical results demonstrate that the OL with $\kappa =0$ cannot sustain stable solitons in the case $\delta=\pi$ either.
For that reason, the analysis reported below focuses only on the case of nonzero $\kappa$, i.e., $\kappa =1$
fixed by scaling in Eq. (\ref{General2DExternalPotential}).
As said above, the choice of $\kappa $ and $\varepsilon
$ as per Eq. (\ref{2}) implies periodic alternation between the OL with the
largest depth and free space, with no OL potential.

\subsection{The Gross-Pitaevskii equation with self-attraction}

\label{sec:SelfAttractiveSymmetricModulation}

\subsubsection{Variational results}

\label{sec:SelfAttractiveSymVA}

For the attractive nonlinearity, with $\sigma =-1$, we first investigated
stability of the fundamental solitons, as predicted by the VA, by
numerically simulating Eq. (\ref{WidthVaEquations}) for the evolution of
widths $W_{x}$ and $W_{y}$. Several examples for the so generated stability
diagram in the $(\omega ,\varepsilon )$ plane are plotted in Figs.~\ref%
{VaStabilityDiagram_1PSin} and \ref{VaStabilityDiagram_1PSin_Exact}, for
inputs with:

\begin{equation}
W_{x}=W_{y}=0.35,N=6.2,  \label{W0p35N6p2}
\end{equation}%
\begin{equation}
W_{x}=W_{y}=0.35,N=5.8,  \label{W0p35N5p8}
\end{equation}%
\begin{equation}
W_{x}=W_{y}=0.25,N=3.0,  \label{W0p25N3p0}
\end{equation}%
\begin{equation}
W_{x}=W_{y}=0.22,N=5.29,  \label{W0p22N5p29}
\end{equation}%
\begin{equation}
W_{x}=W_{y}=0.43,N=2.087,  \label{W0p43N2p087}
\end{equation}%
which correspond to panels (a), (b) and (c) in Fig.~\ref%
{VaStabilityDiagram_1PSin} and panels (a) and (b) in Fig.~\ref%
{VaStabilityDiagram_1PSin_Exact}, respectively. As explained below, the
Gaussian \emph{ans\"{a}tze} with parameters (\ref{W0p22N5p29}) and (\ref%
{W0p43N2p087}) are close to numerically exact stationary solutions found in
the static OL potential, given by Eq. (\ref{Basic2DExternalPotential}), with
$\varepsilon =0$ and the following sets of values of the strength of
the static lattice potential and soliton's chemical potential:
\begin{equation}
{V_{0}=5,\mu =-15},  \label{V05Mum15}
\end{equation}%
\begin{equation}
{V_{0}=5,\mu =-7.4},  \label{V05Mum7p4}
\end{equation}%
respectively. Both solitons belong to the semi-infinite gap in the linear
spectrum induced by the static potential.

\begin{figure}[tbp]
\centering
\noindent\makebox[\textwidth]{
\subfigure[]{\includegraphics[width=3.2in]{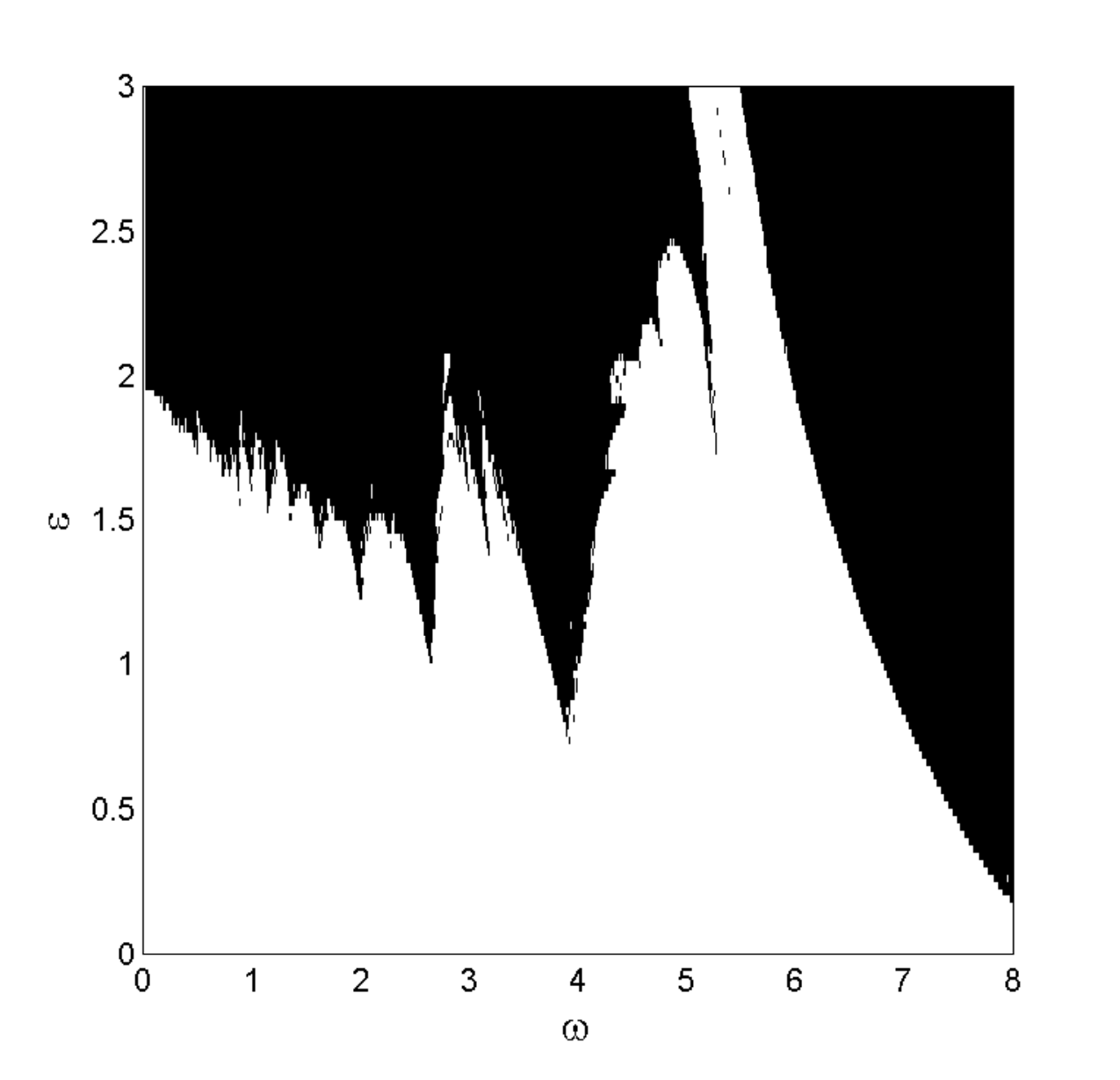}
\label{VaStabilityDiagram_1PSin_Wxy0p35N6p2}}
\subfigure[]{\includegraphics[width=3.2in]{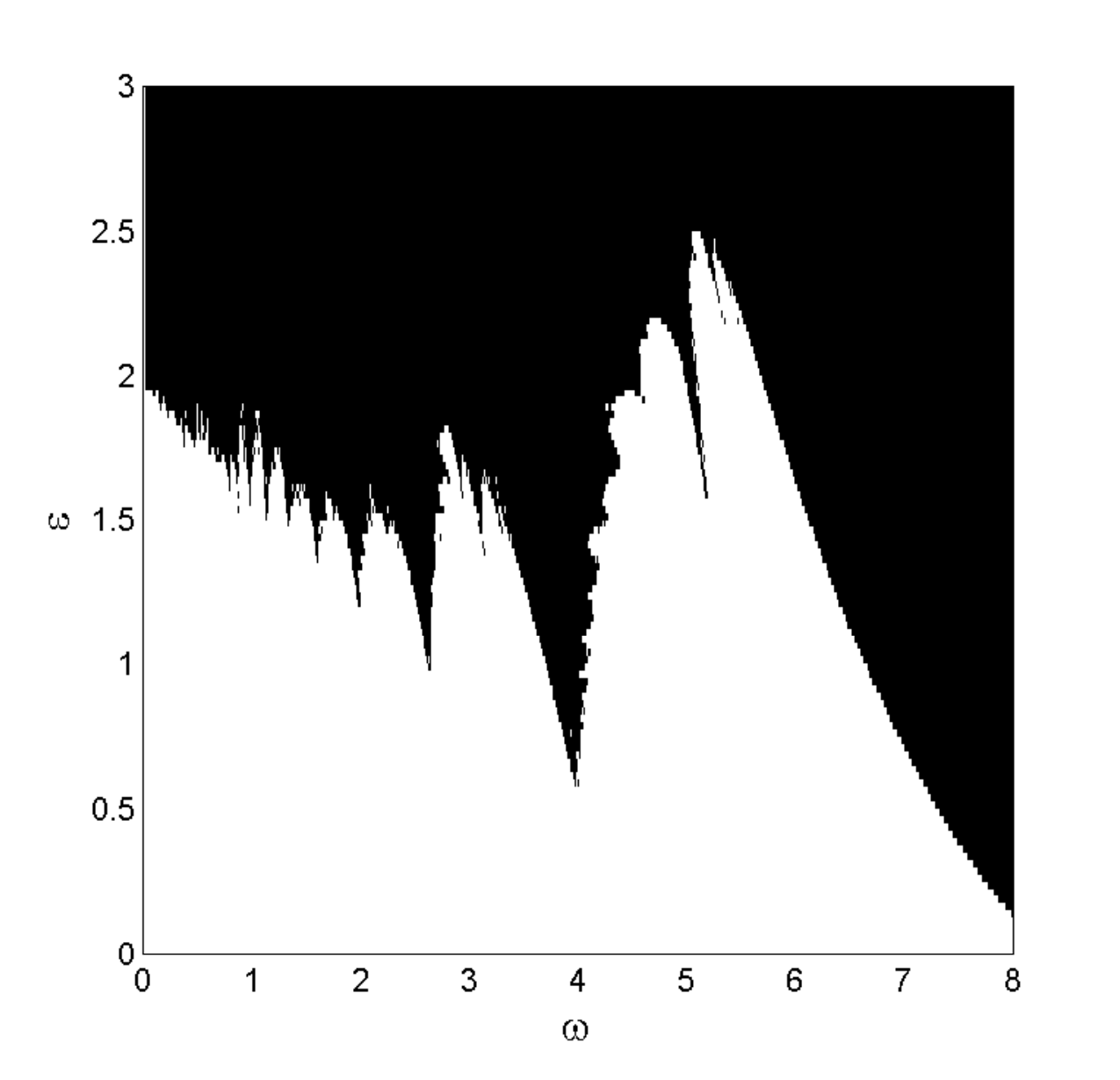}}
\label{VaStabilityDiagram_1PSin_Wxy0p35N5p8}}
\subfigure[]{\includegraphics[width=3.2in]{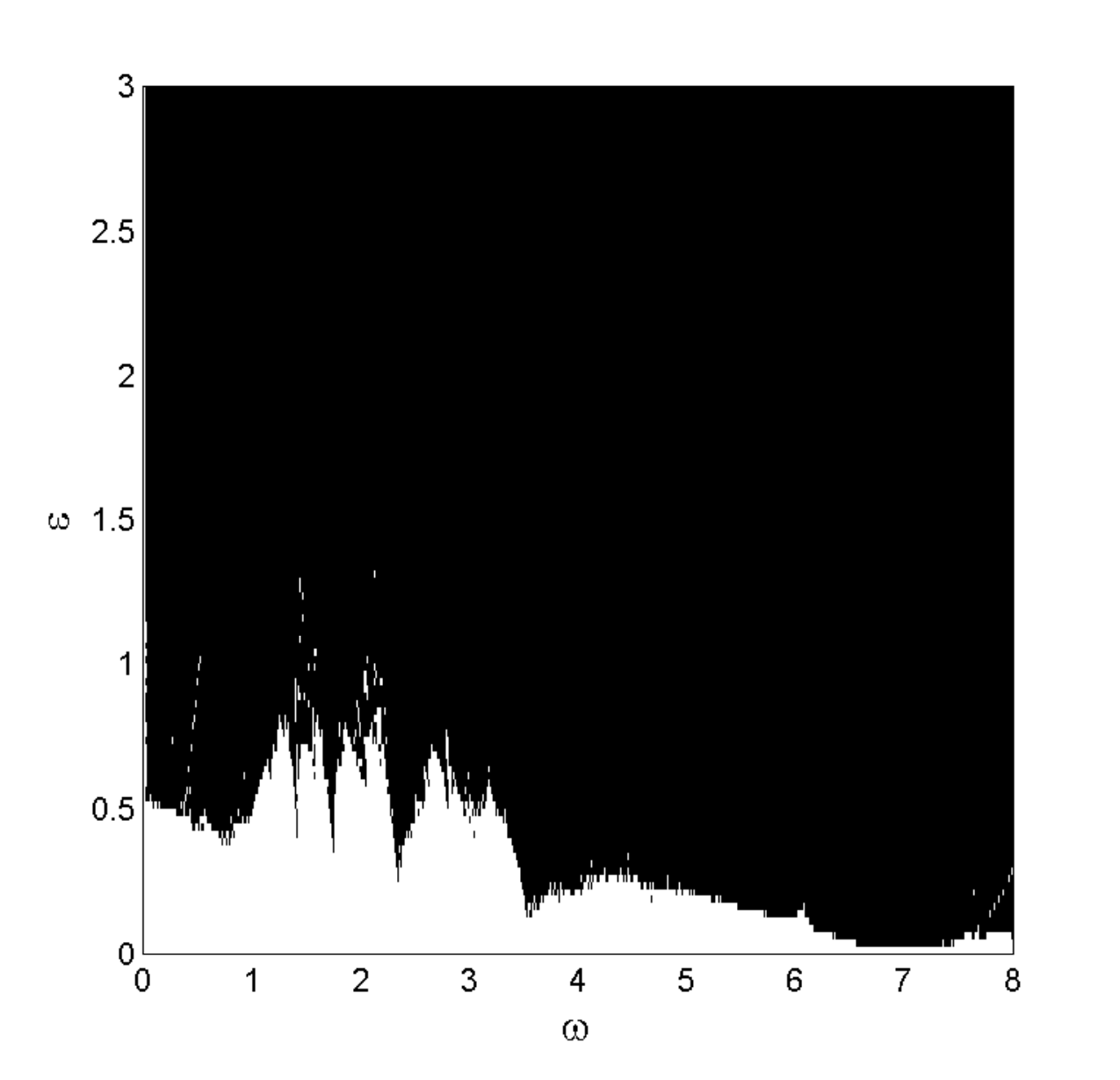}
\label{VaStabilityDiagram_1PSin_Wxy0p25N3p0}}
\caption{Typical stability diagrams for fundamental 2D solitons under the
action of the synchronous time modulation [$\protect\delta =0$ in Eqs. (%
\protect\ref{General2DExternalPotential})], in the case of the attractive
nonlinearity, $\protect\sigma =-1$, in the $(\protect\omega ,\protect%
\varepsilon )$ plane, produced by simulations of the variational evolution
equations for the widths, Eqs. (\protect\ref{WidthVaEquations}). Initial
conditions (\protect\ref{W0p35N6p2}), (\protect\ref{W0p35N5p8}) and (\protect
\ref{W0p25N3p0}), correspond, severally, to panels (a), (b) and (c). Here
and in other figures, white and black regions refer to stable and unstable
solutions, respectively.}
\label{VaStabilityDiagram_1PSin}
\end{figure}

\begin{figure}[tbp]
\centering
\noindent\makebox[\textwidth]{
\subfigure[]{\includegraphics[width=3.2in]{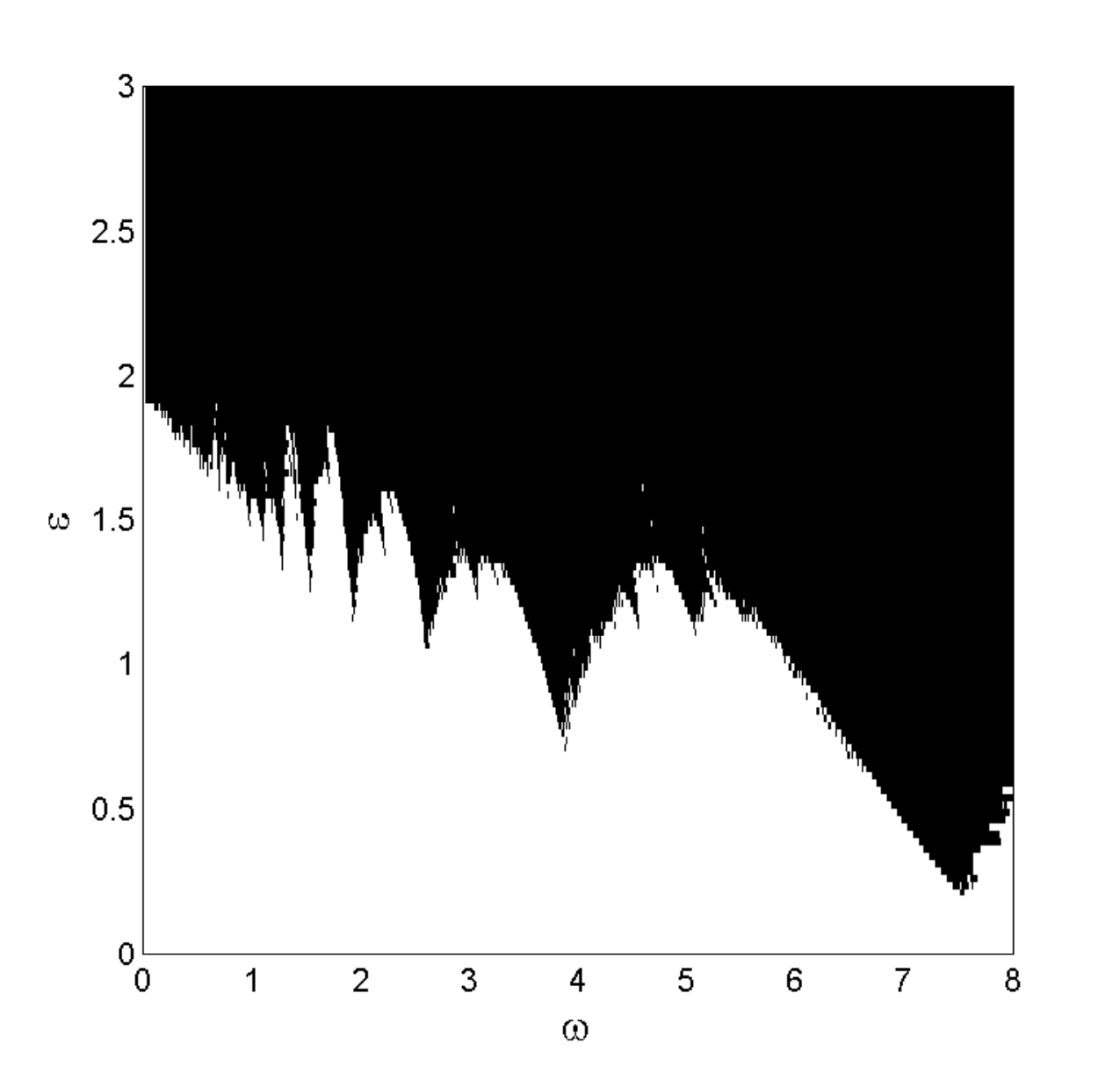}
\label{VaStabilityDiagram_1PSin_Wxy0p22N5p29}}
\subfigure[]{\includegraphics[width=3.2in]{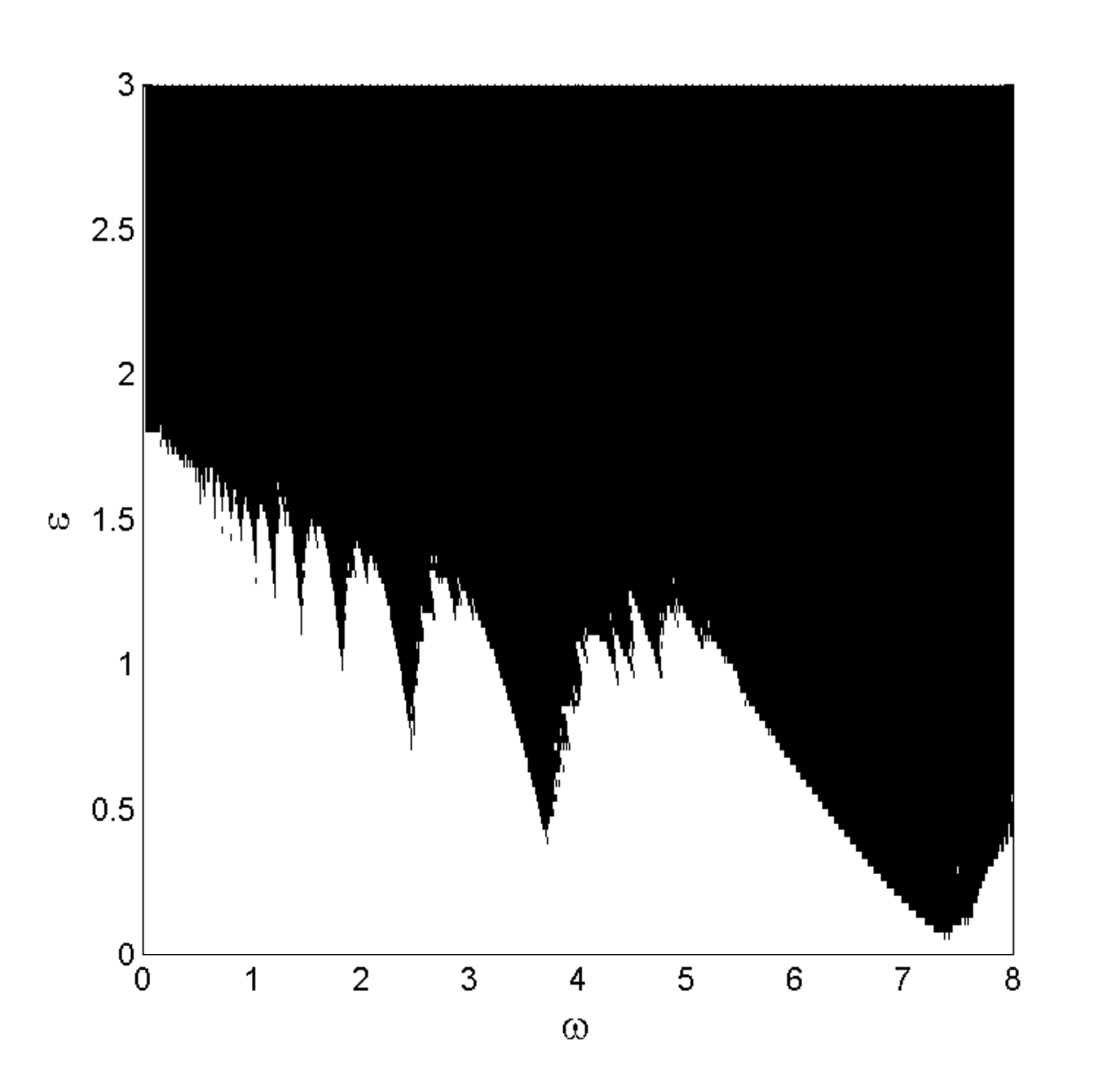}
\label{VaStabilityDiagram_1PSin_Wxy0p43N2p087}}}
\caption{The same as in Fig. ~\protect\ref{VaStabilityDiagram_1PSin}, but
for initial conditions (\protect\ref{W0p22N5p29}) and (\protect\ref%
{W0p43N2p087}), which correspond to panels (a) and (b), respectively.}
\label{VaStabilityDiagram_1PSin_Exact}
\end{figure}

Naturally, the instability appears if the modulation amplitude, $\varepsilon
$, is large enough. Apparent differences between the diagrams in Figs.~\ref%
{VaStabilityDiagram_1PSin} and \ref{VaStabilityDiagram_1PSin_Exact}
demonstrate strong dependence of the stability picture on initial values of
parameters of the Gaussian, $W_{x,y}$ and $N$. Systematic simulations of the
variational dynamical equations (\ref{WidthVaEquations}) show that these
parameters can be chosen to produce optimized stability schemes, with
expanded stable regions and deep and distinctive tongues of instability,
which allow easy identification of resonance frequencies (as elaborated
below).

An important conclusion is that, as the norm of the Gaussian increases,
stability peaks grow in the diagrams, given that the width is properly
chosen (following a criterion mentioned below). This feature can be seen
right away, comparing panels (a) and (b) in Fig.~\ref%
{VaStabilityDiagram_1PSin}, which are produced for the same width, and
different norms, $N=6.2$ and $5.8$. Moreover, it was found that, depending
on the width used, a vast range of values of the norm give rise to a
well-structured stability pattern, featuring increasingly broadened
well-defined stable regions, separated by unstable tongues, see Figs. \ref%
{VaStabilityDiagram_1PSin}(a,b) and \ref{VaStabilityDiagram_1PSin_Exact}%
(a,b). On the other hand, the stability diagram in Fig. \ref%
{VaStabilityDiagram_1PSin_Wxy0p25N3p0}, produced for parameters (\ref%
{W0p25N3p0}), demonstrate the opposite scenario, where the stability regions
are significantly suppressed and distorted. To determine, at least
approximately, ranges of parameters of the initial Gaussian for which the
stability diagrams feature identifiable stable peaks, we consider a
particular reference point in the $(\omega ,\varepsilon )$ plane, chosen
following a systematic analysis which has revealed that the corresponding
initial pulse is stable in well-structured stability areas and unstable
elsewhere. By fixing the modulation parameters corresponding to this
reference point, and testing the predicted stability as a function of the
Gaussian's norm and width, the corresponding stability chart, in the $%
(W_{x,y},N)$ plane, can be constructed, as shown in Fig.~\ref%
{VaStabilityDiagram_1PSin_Omega5Epsilon1} for a particular reference point,

\begin{equation}
{\omega=5, \varepsilon=1},  \label{RefPointOmega5Epsilon1}
\end{equation}

\begin{figure}[tbp]
\centering
\subfigure[]{\includegraphics[width=3.2in]{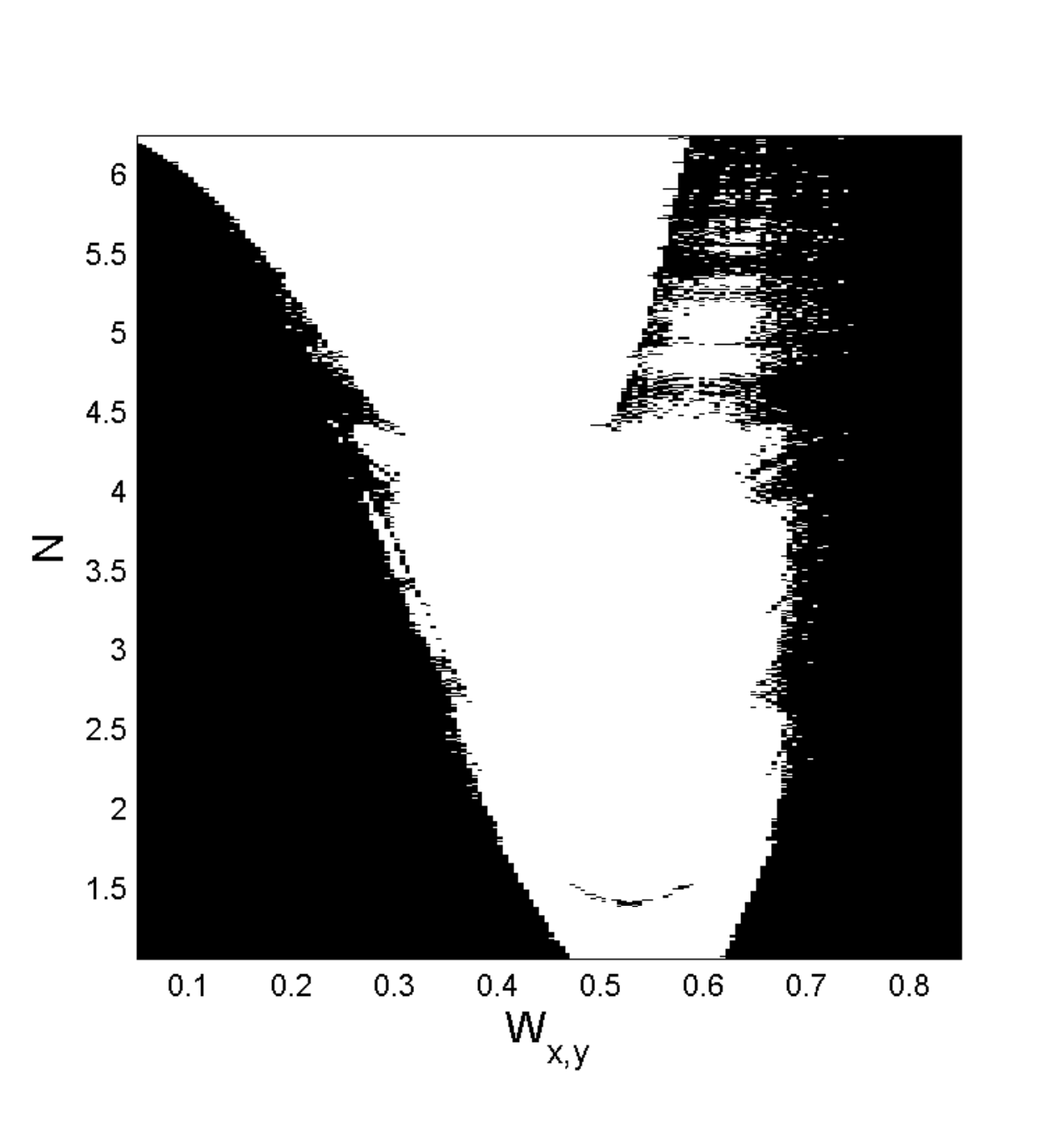}
\label{VaStabilityDiagram_1PSin_Omega5Epsilon1}}
\caption{The stability area (white), obtained via simulations of variational
equations (\protect\ref{WidthVaEquations}), in the plane of parameters of
the Gaussian ansatz, $(W_{x}=W_{y};N)$, for synchronous time modulation ($%
\protect\delta =0$) and self-attractive nonlinearity ($\protect\sigma =-1$).
The parameters of the temporal modulation are fixed as $\protect\omega =5$
and $\protect\varepsilon =1$.}
\label{VaStabilityDiagram_1PSin_Omega5Epsilon1}
\end{figure}

Examination of the stability/instability domains in the diagram displayed in
Fig. \ref{VaStabilityDiagram_1PSin_Omega5Epsilon1} demonstrates that optimal
stability patterns are achieved when taking the initial pulse's width to be
close to the center of the displayed stability region, and considering large
values of the norm. The sets of initial values in Eqs. (\ref{W0p35N6p2}) and
(\ref{W0p35N5p8}) are typical examples of such values which generate broad
stability areas. This procedure was also implemented for other
time-modulation scenarios elaborated below, for identifying optimal
stability patterns.

When focusing on the optimized stability diagram in Fig.~\ref%
{VaStabilityDiagram_1PSin_Wxy0p35N6p2}, for the initial values of the
Gaussian given in Eq. (\ref{W0p35N6p2}), one can clearly identify \emph{%
instability tongues}, which are generated by resonances at the following
values of the driving frequency (for ten largest tongues in the
frequency-domain investigated in the present work, $\omega <20$):
\begin{equation}
\omega _{0}^{\mathrm{(VA)}}=16.25,8.05,3.9,2.65,2.0,1.62,1.35,1.14,0.98,0.89.
\label{res-VA}
\end{equation}%
Similar resonant values, different from those in Eq. (\ref{res-VA}) by less
than $1\%$, are also found in the second optimized stability diagram in Fig.~%
\ref{VaStabilityDiagram_1PSin_Wxy0p35N5p8}, for the initial values chosen as
per Eq. (\ref{W0p35N5p8}). For the two stability diagrams in Fig.~\ref%
{VaStabilityDiagram_1PSin_Exact}(a,b), which corresponds to points
positioned at the edge of the stability domain of Fig.~\ref%
{VaStabilityDiagram_1PSin_Omega5Epsilon1}, the resonance frequencies are
also relatively close to those in (\ref{res-VA}), with a difference $<5\%$.

As demonstrated in diagrams \ref{VaStabilityDiagram_1PSin}(a,b), \ref%
{VaStabilityDiagram_1PSin_Exact}(a,b) and in similar stability diagrams
displayed below, as the resonant frequencies decrease, their detection
becomes increasingly more difficult. This feature is attributed to the
corresponding decrease of the instability growth rate, to the point where
the instability does not develop in course of the finite simulation time,
and the resonant frequencies can no longer be distinguished. For this
reason, the instability tongues that correspond to the lower resonant
frequencies in (\ref{res-VA}), shrink and terminate at finite values of $%
\varepsilon $.

When exploring the stability for modulation frequencies larger than those
corresponding to Figs.~\ref{VaStabilityDiagram_1PSin} and \ref%
{VaStabilityDiagram_1PSin_Exact}, an additional stability peak is found in
the range of $8\lesssim \omega \lesssim 16$, higher than the previous one,
extending beyond the critical value (\ref{2}), for sets (\ref{W0p35N6p2}), (%
\ref{W0p35N5p8}), (\ref{W0p22N5p29}) and (\ref{W0p43N2p087}).
This outcome is expected for cases corresponding to Eqs. (\ref{W0p35N6p2})
and (\ref{W0p35N5p8}), when the stability peaks, shown in Figs.~\ref%
{VaStabilityDiagram_1PSin_Wxy0p35N6p2} and \ref%
{VaStabilityDiagram_1PSin_Wxy0p35N5p8}, are steadily growing with the
increase of $\omega $. For the parameters corresponding to Eqs. (\ref%
{W0p22N5p29}) and (\ref{W0p43N2p087}), this result is more surprising, as
all the preceding stability peaks observed in Figs.~\ref%
{VaStabilityDiagram_1PSin_Exact}(a,b) are systematically decreasing.
Frequencies $\omega >20$ are not considered here.

\subsubsection{Numerical results}
\label{sec:SelfAttractiveSymNum}
The analysis of the variational system was followed by systematic
simulations of GPE (\ref{2DGPE}). Here we produce findings concerning the
stability for the initial conditions taken in the form of solitons created
as stationary solutions of the static version of Eq. (\ref{2DGPE}), with $%
\varepsilon =0$, at different points in the semi-infinite gap, with the same
time-modulation parameters as used above. The GPE simulations were run up to
$t=1000$. The resulting dynamical soliton was registered as a stable one if
it kept, at $t=1000$, $\geq 99\%$ of the initial norm (the rest might be
lost through emission of radiation). First we address the stability of a
soliton positioned relatively close to the edge of the semi-infinite gap,
corresponding to parameter values (\ref{V05Mum7p4}). The respective
stability diagram is displayed in Fig.~\ref%
{StabilityDiagram_1PSin_Wxy0p43N2p087}.

\begin{figure}[tbp]
\centering
\noindent\makebox[\textwidth]{
\subfigure[]{\includegraphics[width=3.2in]{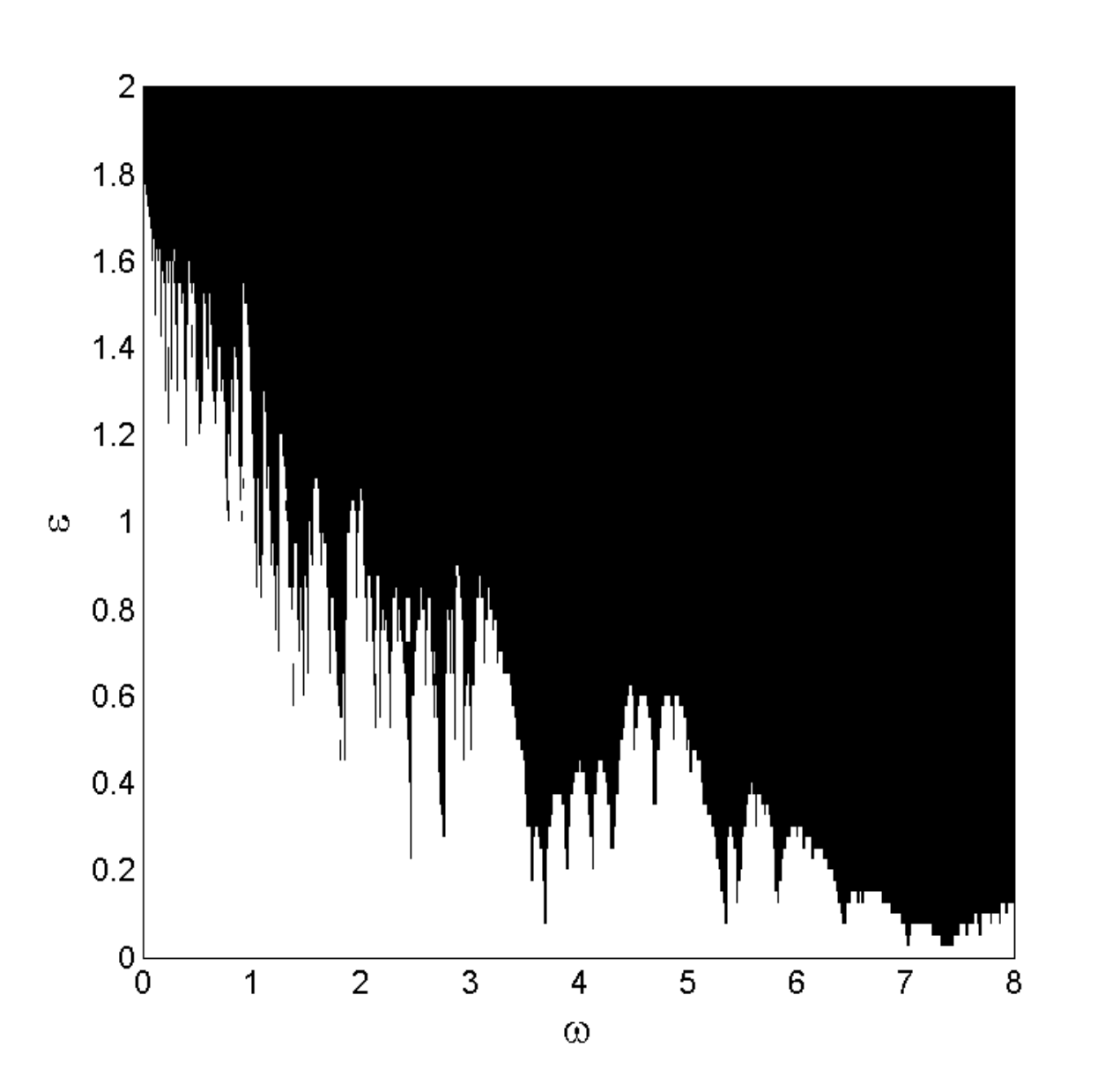}
\label{StabilityDiagram_1PSin_Wxy0p43N2p087}}
\subfigure[]{\includegraphics[width=3.2in]{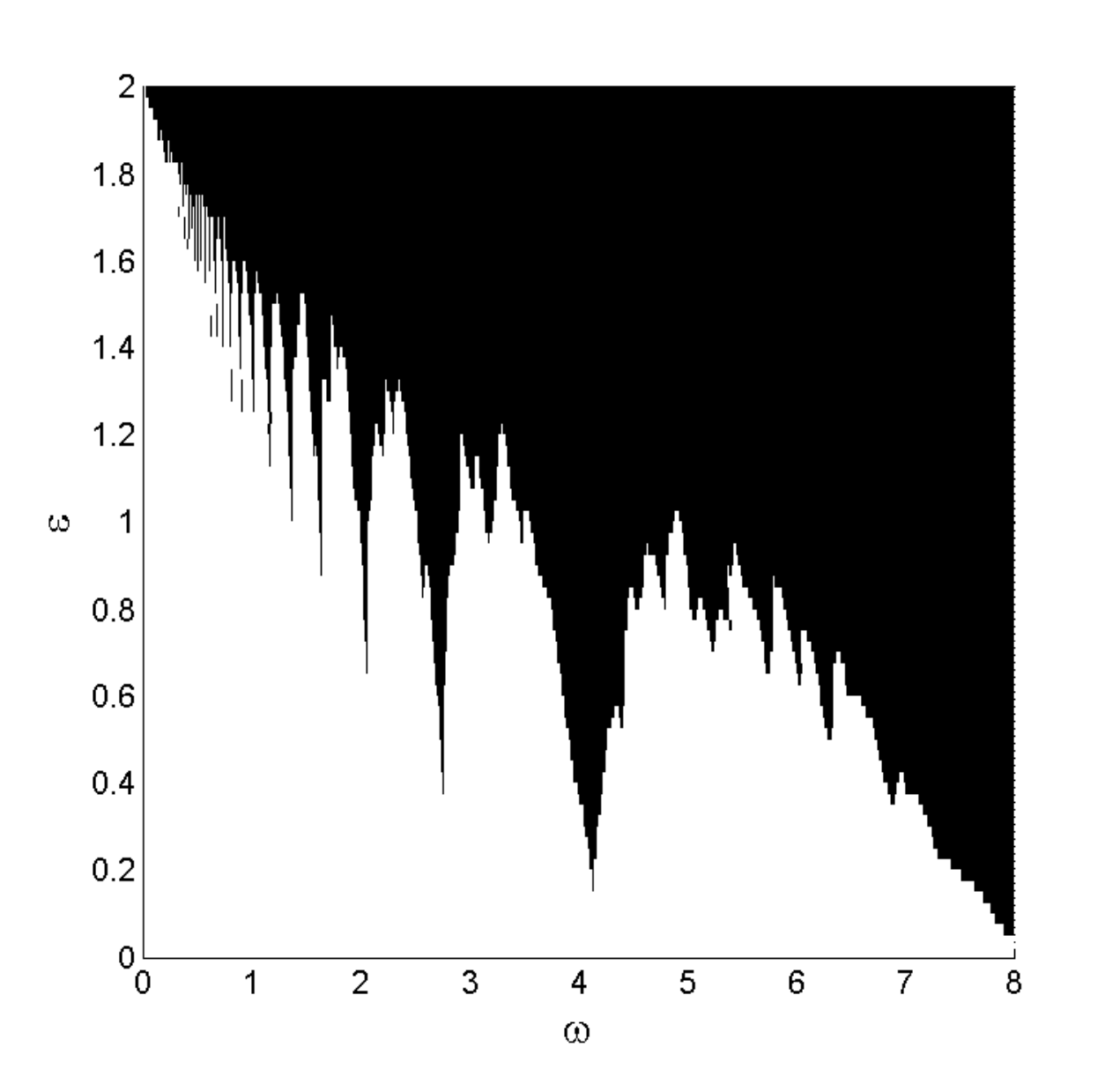}
\label{StabilityDiagram_1PSin_Wxy0p22N5p29}}}
\caption{Stability diagrams for the fundamental 2D solitons, in the $(%
\protect\omega ,\protect\varepsilon )$ plane, as found from direct
simulations of Eq. (\protect\ref{2DGPE}) with the self-attractive
nonlinearity, $\protect\sigma =-1$, and synchronous time modulation ($%
\protect\delta =0$) applied to the OL. The inputs were taken as the
numerically exact stationary soliton solutions of the static version of the
equation ($\protect\varepsilon =0$) at points corresponding to Eqs. (\protect
\ref{V05Mum7p4}) and (\protect\ref{V05Mum15}), in panels (a) and (b),
respectively.}
\label{StabilityDiagram_1PSin}
\end{figure}

The so produced stability pattern seems irregular, in comparison with the
more regular one, predicted by the VA for the same parameters in Fig.~\ref%
{VaStabilityDiagram_1PSin_Exact}(b). However, deeper in the semi-infinite
gap, the numerically generated stability diagram becomes more structured,
clearly exhibiting stability peaks and instability tongues. An example is
shown in Fig.~\ref{StabilityDiagram_1PSin_Wxy0p22N5p29}, for parameter
values (\ref{V05Mum15}). In particular, this diagram features a set of
instability tongues at the following resonant frequencies:
\begin{equation}
\omega _{0}^{\mathrm{(num)}%
}=15.94,8.19,4.13,2.75,2.06,1.64,1.36,1.16,1.02,0.9,  \label{res-num}
\end{equation}%
which are reasonably close to their VA-predicted counterparts given by Eq. (%
\ref{res-VA}). Nevertheless, further comparison between the stability
diagrams produced by means of the GPE simulations, Fig.~\ref%
{StabilityDiagram_1PSin_Wxy0p22N5p29}, and the corresponding VA prediction
in Fig.~\ref{VaStabilityDiagram_1PSin_Wxy0p22N5p29}, shows significant
differences with respect to the modulation strength, as the VA predicts
persistence of stability at considerably higher values of $\varepsilon $.
For instance, according to the predicted results, the last stability peak ($%
4\lesssim \omega \lesssim 8$) seen in Fig.~\ref%
{VaStabilityDiagram_1PSin_Wxy0p22N5p29} reaches $\varepsilon _{\max }^{(%
\mathrm{VA})}\cong 1.45$, while the GPE simulations yield $\varepsilon
_{\max }^{(\mathrm{num})}\cong 1.0$.

The stability was also studied for modulation frequencies larger than $%
\omega =8$ [not shown in Fig.~\ref{StabilityDiagram_1PSin_Wxy0p22N5p29}]. In
particular, the GPE simulations have revealed an additional stability
region, bounded by two instability tongues at $8.19<\omega <15.94$, similar
to the predicted one outlined above. This stability region is much wider and
higher than the preceding one, with the top almost reaching the critical
value $\varepsilon _{\mathrm{cr}}$, given by Eq. (\ref{2}). Again, in terms
of the modulation-strength values, this result emphasizes the difference in
comparison with the VA-predicted outcome, where the similar peak extends
well beyond $\varepsilon _{\mathrm{cr}}$.

It should be stressed that, for all the modulation frequencies which we
examined in the present setting (specifically, $0<\omega <20$), the soliton
is always unstable at $\varepsilon \geq \varepsilon _{\mathrm{cr}}$. In
other words, the increase of the modulation frequency cannot compensate for
the destructive effect of the periodically vanishing OL in the modulation
format which implies the periodic switching between the OL and free space,
in the case when Eq. (\ref{2}) holds. As demonstrated below, different
modulation formats, which do not periodically switch off the OL, are able to
support stable solitons at $\varepsilon \geq \varepsilon _{\mathrm{cr}}$.

Additional analysis was also performed for a soliton positioned even deeper
in the semi-infinite gap, with a norm close to the maximal one, $N_{\max
}=5.85$ (the norm of the \emph{Townes} soliton \cite{Berge,Fibich}). For
this purpose, we used the input provided by the numerically exact solution
obtained at $V_{0}=5,\mu =-30$, with norm $N=5.8$. The conclusion is that
the stability map in this case (not shown here) is very similar to the one
displayed for the case of Eq. (\ref{V05Mum15}) in Fig.~\ref%
{StabilityDiagram_1PSin_Wxy0p22N5p29}. In particular, the increase of the
norm did not lead to expansion of the stable regions, because, as a matter
of fact, the stability pattern presented in Fig.~\ref%
{StabilityDiagram_1PSin_Wxy0p22N5p29} is already quite close to the optimal
one.

\subsection{The Gross-Pitaevskii equation with self-repulsion}

\subsubsection{Variational results}

\begin{figure}[tbp]
\centering
\noindent\makebox[\textwidth]{
\subfigure[]{\includegraphics[width=3.2in]{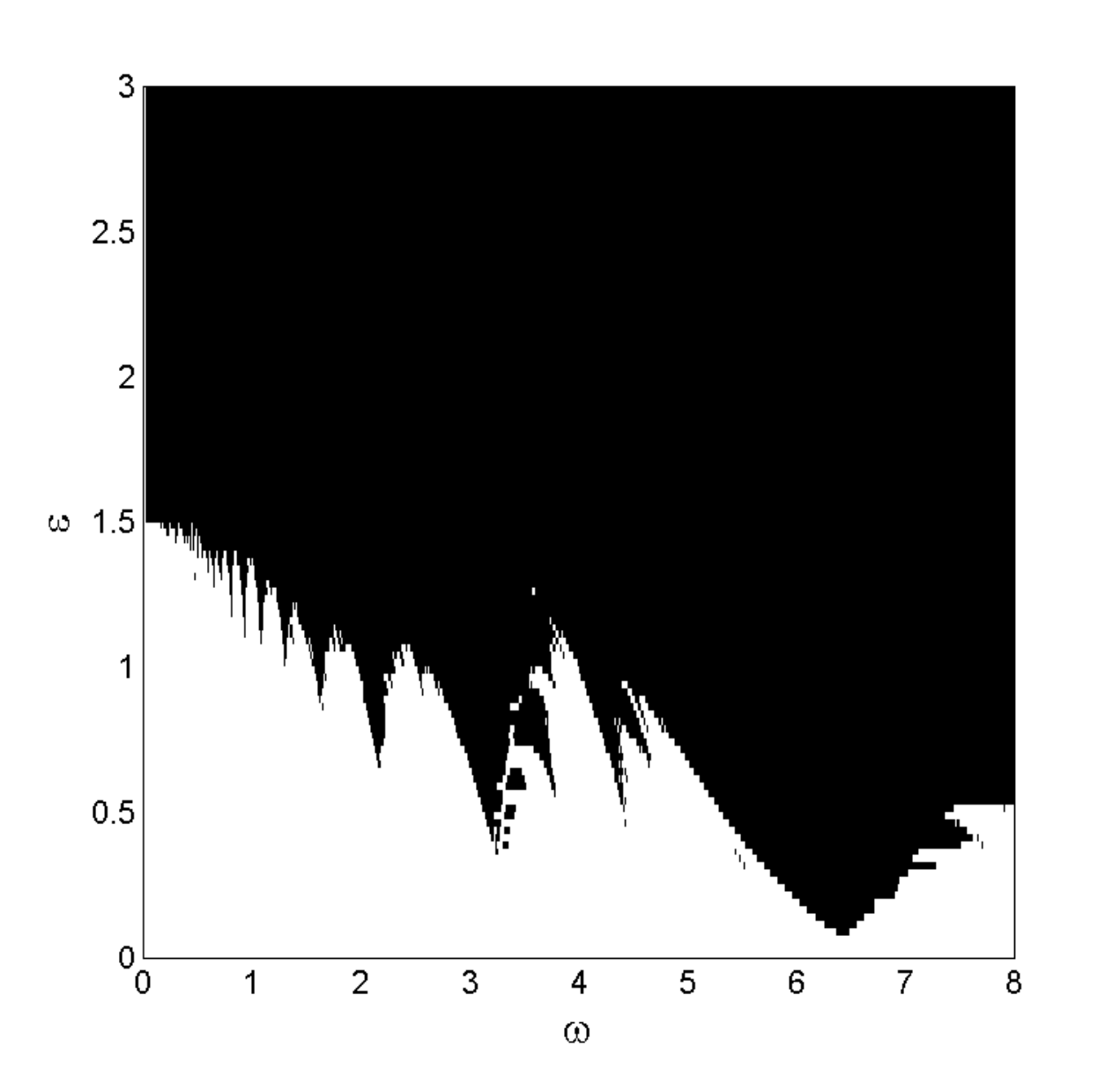}
\label{VaStabilityDiagram_1PSin_Wxy0p615N4p73}}
\subfigure[]{\includegraphics[width=3.2in]{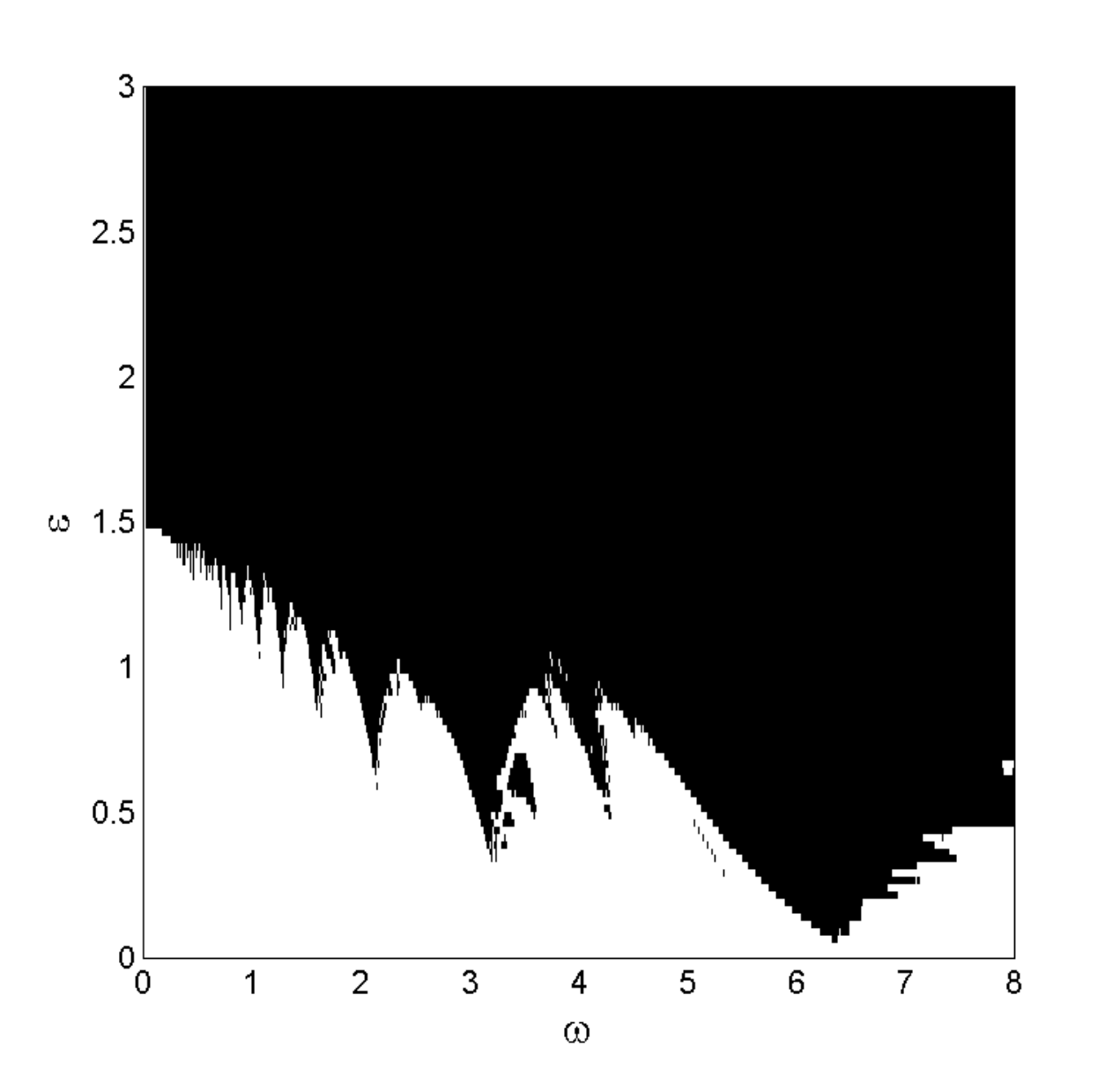}}
\label{VaStabilityDiagram_1PSin_Wxy0p615N6p0}}
\subfigure[]{\includegraphics[width=3.2in]{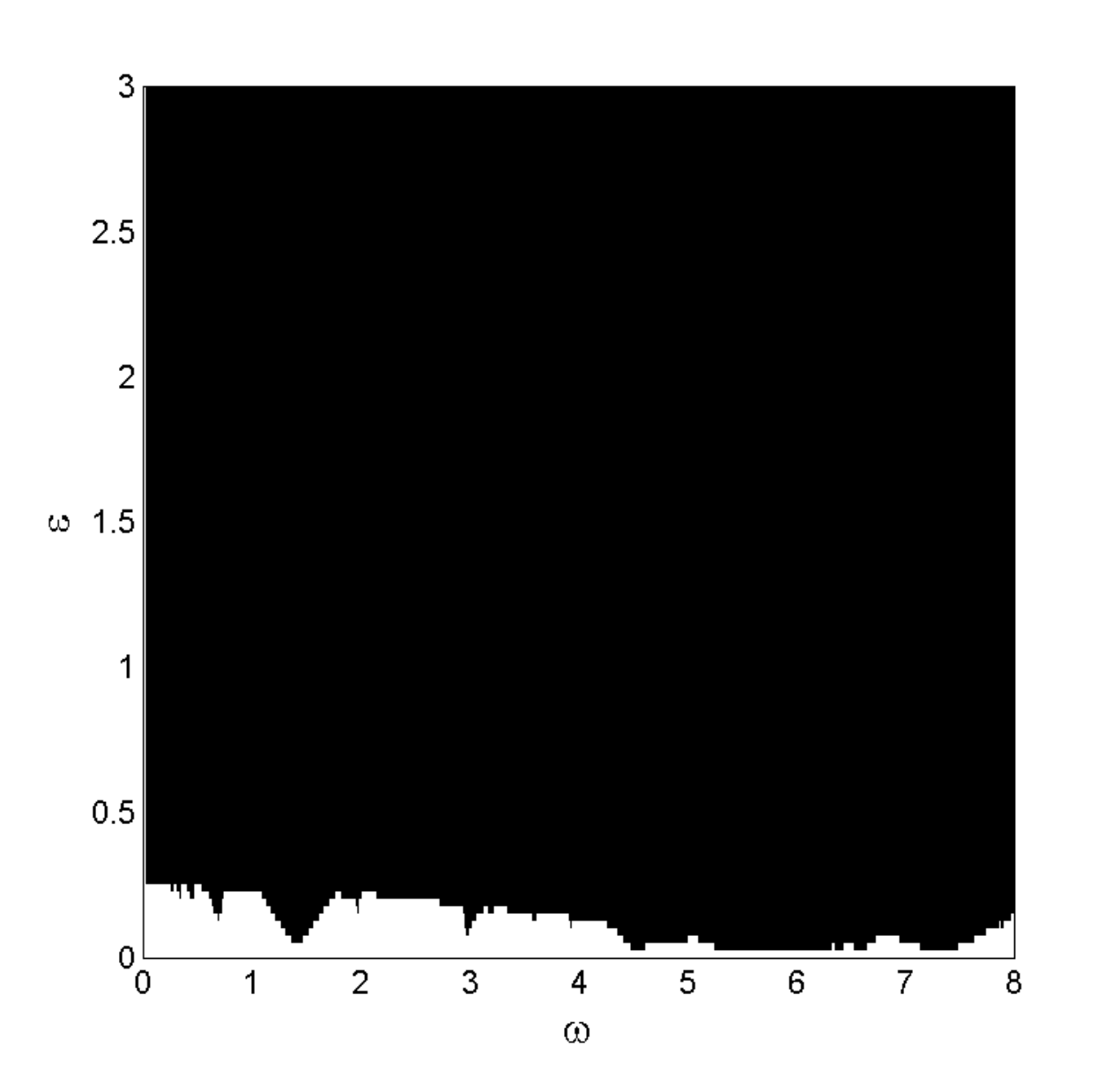}
\label{VaStabilityDiagram_1PSin_Wxy0p4N4p73}}
\caption{The stability diagram for the solitons in the model with the
repulsive nonlinearity and the OL subject to the synchronous modulation ($%
\protect\delta =0$), plotted in the $(\protect\omega ,\protect\varepsilon )$
plane, as produced by the VA. Panels (a), (b) and (c) refer to initial
conditions (\protect\ref{W0p615N4p73}), (\protect\ref{W0p615N6p0}), and (%
\protect\ref{W0p4N4p73}) respectively.}
\label{VaStabilityDiagramRepulsive_1PSin}
\end{figure}

The application of the VA, based on Eq. (\ref{WidthVaEquations}), to the
repulsive model, with $\sigma =+1$ in Eq. (\ref{Confined2DPeriodicPotential}%
), was systematically carried out for a wide range of the parameters of the
initial Gaussian. Figure~\ref{VaStabilityDiagramRepulsive_1PSin} displays
three examples of resulting stability maps in the $(\omega ,\varepsilon )$
plane, for the following sets of initial conditions:
\begin{equation}
W_{x}=W_{y}=0.615,N=4.73,  \label{W0p615N4p73}
\end{equation}%
\begin{equation}
W_{x}=W_{y}=0.615,N=6.0,  \label{W0p615N6p0}
\end{equation}%
\begin{equation}
W_{x}=W_{y}=0.4,N=4.73.  \label{W0p4N4p73}
\end{equation}

Similar to the case of the attractive nonlinearity considered above, the set
corresponding to Eq. (\ref{W0p615N4p73}) was specifically chosen to closely
mimic the numerically exact stable stationary solution found near the middle
of the first finite bandgap of the static OL ($\varepsilon =0$), with
parameters
\begin{equation}
{V_{0}=5,\mu =-3.5}.  \label{V05Mum3p5}
\end{equation}

As seen in Fig.~\ref{VaStabilityDiagramRepulsive_1PSin}, the stability
diagrams for parameters (\ref{W0p615N4p73}) and (\ref{W0p615N6p0}), which
refer to points with the same width but different norms, bear a close
resemblance to each other and demonstrate stability regions that are large
and clearly bounded. On the other hand, an example for a reduced stability
pattern, with no obvious increasingly widened stable peaks, is shown in Fig.~%
\ref{VaStabilityDiagram_1PSin_Wxy0p4N4p73} for the case corresponding to Eq.
(\ref{W0p4N4p73}). To identify a region in the parameter plane where the
structured stability patterns, such as those seen in panels (a) and (b) of
Fig. \ref{VaStabilityDiagramRepulsive_1PSin}, may be obtained, we followed
the same procedure as in the case of the attractive nonlinearity, and have
thus spotted a point of reference in the plane $(\omega ,\varepsilon )$,
which roughly allows us to differentiate between the two stability
scenarios. Here, we chose the point as $\omega =1.75$ and $\varepsilon =1$
[inside the stable domain in Fig.~\ref{VaStabilityDiagramRepulsive_1PSin}%
(a,b), and outside of it in Fig.~\ref{VaStabilityDiagramRepulsive_1PSin}%
(c)], and constructed the stability chart displayed in Fig.~\ref%
{RepStabilityDiagram_1PSin_Omega1p75}. Similar to the variational prediction
in the case of the attractive nonlinearity, Fig. \ref%
{RepStabilityDiagram_1PSin_Omega1p75} shows that well-structured stability
charts are produced for a wide range of values of the norm, given that the
width of the initial pulse is suitably selected.
\begin{figure}[tbp]
\centering
\subfigure[]{\includegraphics[width=3.4in]{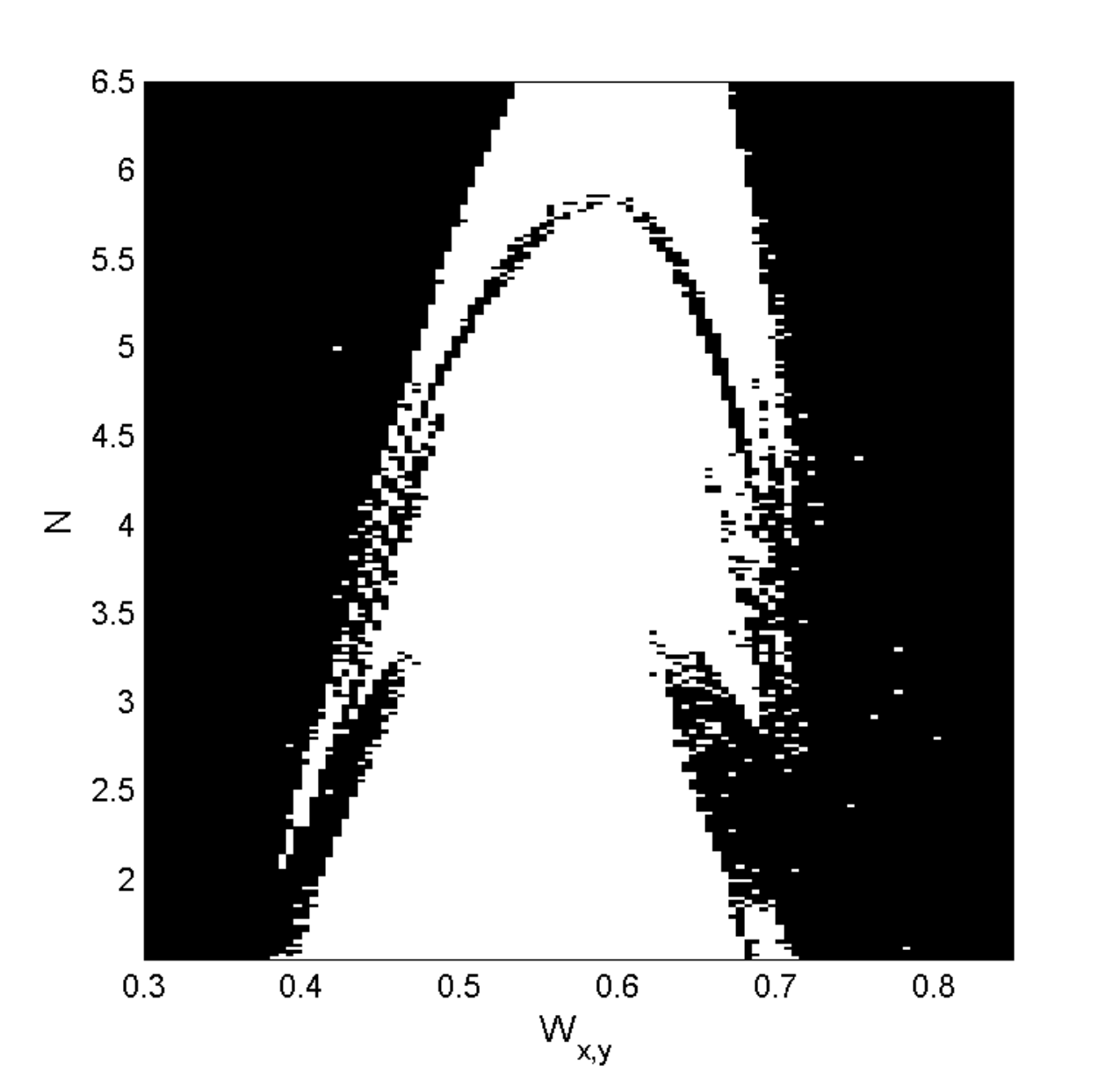}
\label{RepStabilityDiagram_1PSin_Omega1p75}}
\caption{The stability area in the plane of the parameters of the Gaussian
ansatz, $(W_{x}=W_{y},N)$, similar to Fig.~\protect\ref%
{VaStabilityDiagram_1PSin_Omega5Epsilon1}, but for the repulsive
nonlinearity. The modulated OL is taken with $\protect\varepsilon =1$ and $%
\protect\omega =1.75$.}
\label{RepStabilityDiagram_1PSin_Omega1p75}
\end{figure}
Stability profiles of this type feature sets of distinctive instability
tongues, originating, for the two particular examples introduced above, Fig.~%
\ref{VaStabilityDiagramRepulsive_1PSin}(a) and (b), at frequencies:

\begin{equation}
\omega _{0}=11.62,6.40,3.25,2.15,1.63,1.28,1.07,0.91,0.8,0.71.
\label{res-VA-rep}
\end{equation}

\subsubsection{Numerical results}

\begin{figure}[tbp]
\centering
\subfigure[]{\includegraphics[width=3.4in]{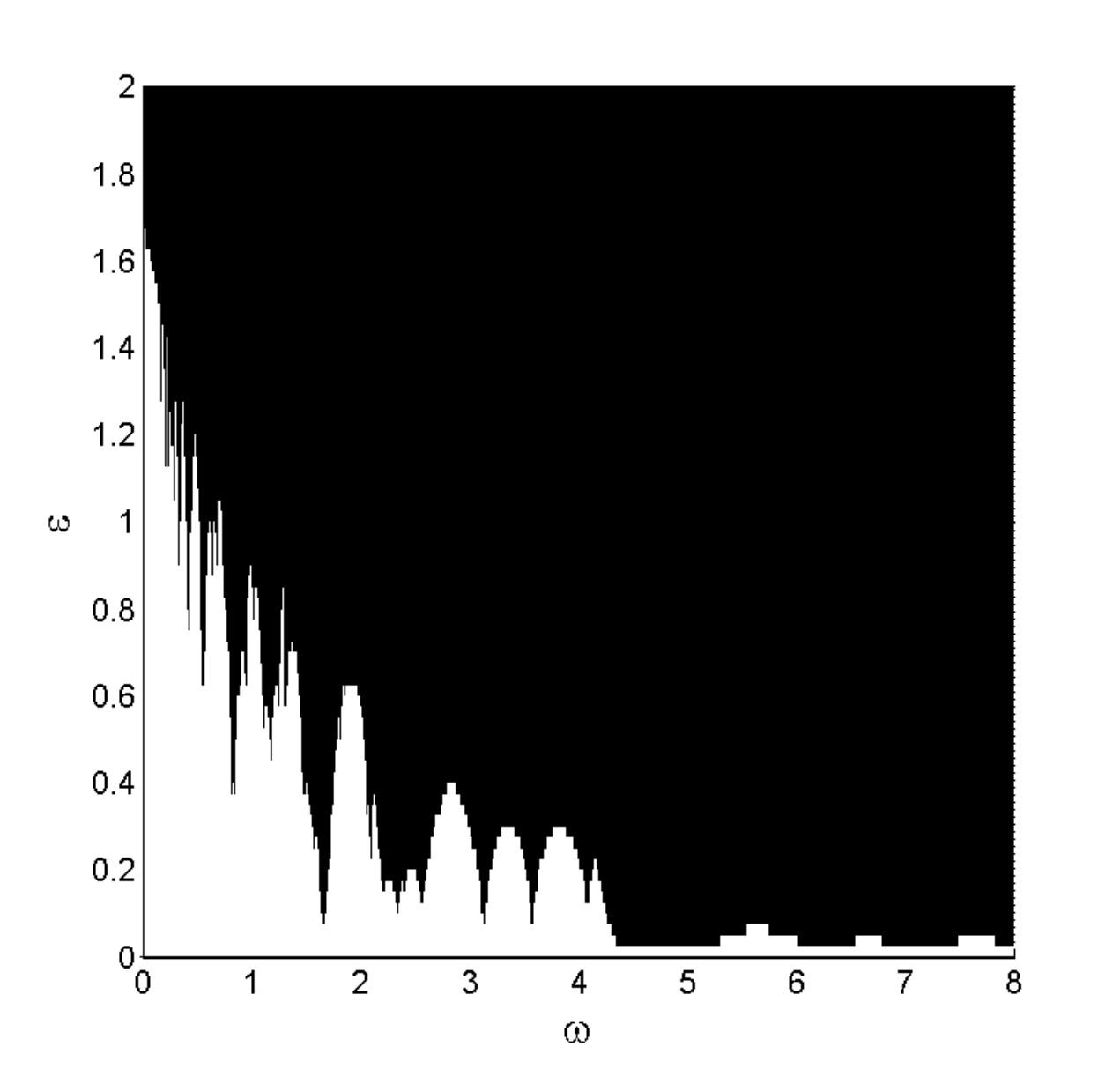}
\label{StabilityDiagram_1PSin_Wxy0p615N4p73}}
\caption{The same as in Fig.~\protect\ref{StabilityDiagram_1PSin}, but for
the repulsive nonlinearity, $\protect\sigma =+1$. Here, the initial
condition is a numerically exact gap soliton with parameters given by Eq. (%
\protect\ref{V05Mum3p5}).}
\label{StabilityDiagramRepulsive_1PSin}
\end{figure}

Direct simulations of the underlying GPE (\ref{2DGPE}) with $\sigma =+1$
have been performed too. The respective stability diagram was produced for
the input taken as a numerically exact gap soliton with parameters (\ref%
{V05Mum3p5}) in the first finite bandgap, supported by static potential (\ref%
{Basic2DExternalPotential}). In the diagram, displayed in Fig.~\ref%
{StabilityDiagramRepulsive_1PSin}, the exact stability profile is quite
different from the VA-predicted one in Fig.~\ref%
{VaStabilityDiagramRepulsive_1PSin}(a). Mainly, the variational stable
regions, with growing width and tidy shapes, are replaced by stability peaks
with irregular changes in their size. In addition, for large frequencies ($%
\omega >4.4$, in the case of Fig.~\ref{StabilityDiagramRepulsive_1PSin}),
the solutions are almost completely unstable, even for low values of the
modulation strength, while the VA does predict considerable stability
regions in the entire frequency domain considered in this work, $\omega >16$
[not fully shown in Figs~\ref{VaStabilityDiagramRepulsive_1PSin}(a,b)].
Point (\ref{V05Mum3p5}) considered above is located near the center of the
first finite bandgap. When moving closer to the edges of the bandgap, the
stationary GSs exhibit increasingly growing tails, which are found to
accelerate the onset of the instability, in the presence of the time
modulation. Thus, stability diagrams for such initial solutions introduce
even smaller stability areas than the one displayed in Fig.~\ref%
{StabilityDiagramRepulsive_1PSin} (not shown here in detail).

\section{The 2D lattice under asynchronous modulation of the 1D sublattices}

\label{sec:NonsynchronousModulation}

\subsection{The phase shift of $\protect\delta =\protect\pi /2$ between the
sublattices}

Here we address the underlying model with phase shift $\delta =\pi /2$
between the modulations applied to the $x$ and $y$ sublattices in Eq. (\ref%
{General2DExternalPotential}).
Stable solitons for the phase-shifted modulation ($\delta =\pi /2$) can be
found in the model which includes the static component of the lattice
potential (\ref{General2DExternalPotential}), with $\kappa =1$ fixed by
scaling, as mentioned above. The analysis in this case was to the
attractive nonlinearity, with $\sigma =-1$, hence the respective initial
conditions were taken as stationary solitons in the semi-infinite bandgap of
the static OL.

\subsubsection{Variational results}
\label{sec:NonsynHalfPhiVA}
First, we consider the same reference point that was selected for the
(attractive) synchronous setting, (\ref{RefPointOmega5Epsilon1}), and
construct the corresponding stability map shown in Fig.~\ref%
{VaStabilityDiagram_SinCos_Omega5Epsilon1}, in the $(W_{x}=W_{y},N)$ plane,
within the framework of the variational equations (\ref{WidthVaEquations}).

\begin{figure}[tbp]
\centering
\subfigure[]{\includegraphics[width=3.4in]{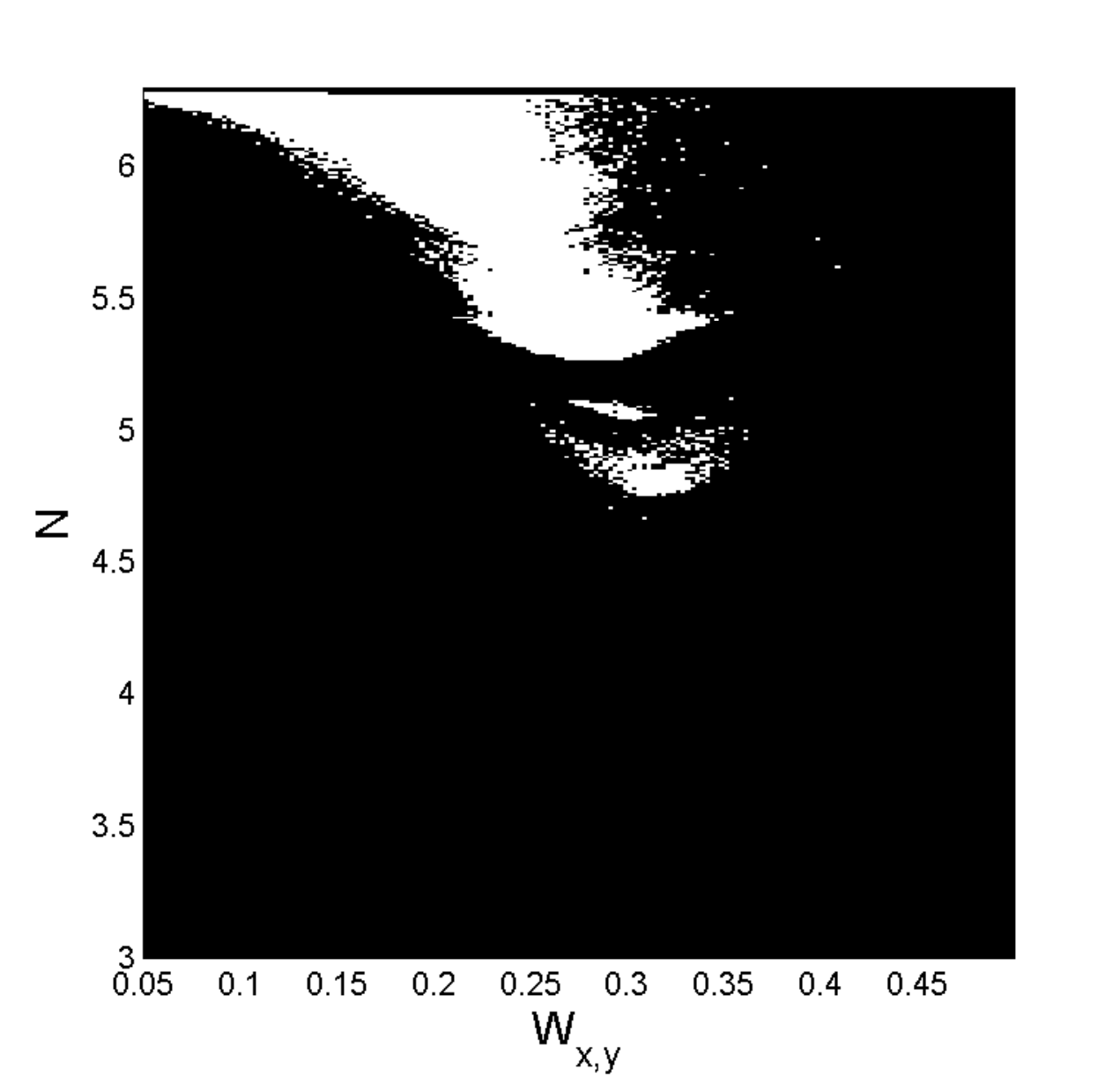}
\label{VaStabilityDiagram_SinCos_Omega5Epsilon1}}
\caption{A stability chart\ in the $\left( W_{x},=W_{y},N\right) $ plane, as
produced by simulations of the variational equations (\protect\ref%
{WidthVaEquations}), for the OL with $\protect\kappa =1$, $%
\protect\delta =\protect\pi /2$, and the time-modulation parameters taken as
per Eq.  (\protect\ref{RefPointOmega5Epsilon1}). }
\label{VaStabilityDiagram_SinCos_Omega5Epsilon1}
\end{figure}

\begin{figure}[tbp]
\centering
\noindent\makebox[\textwidth]{
\subfigure[]{\includegraphics[width=3.4in]{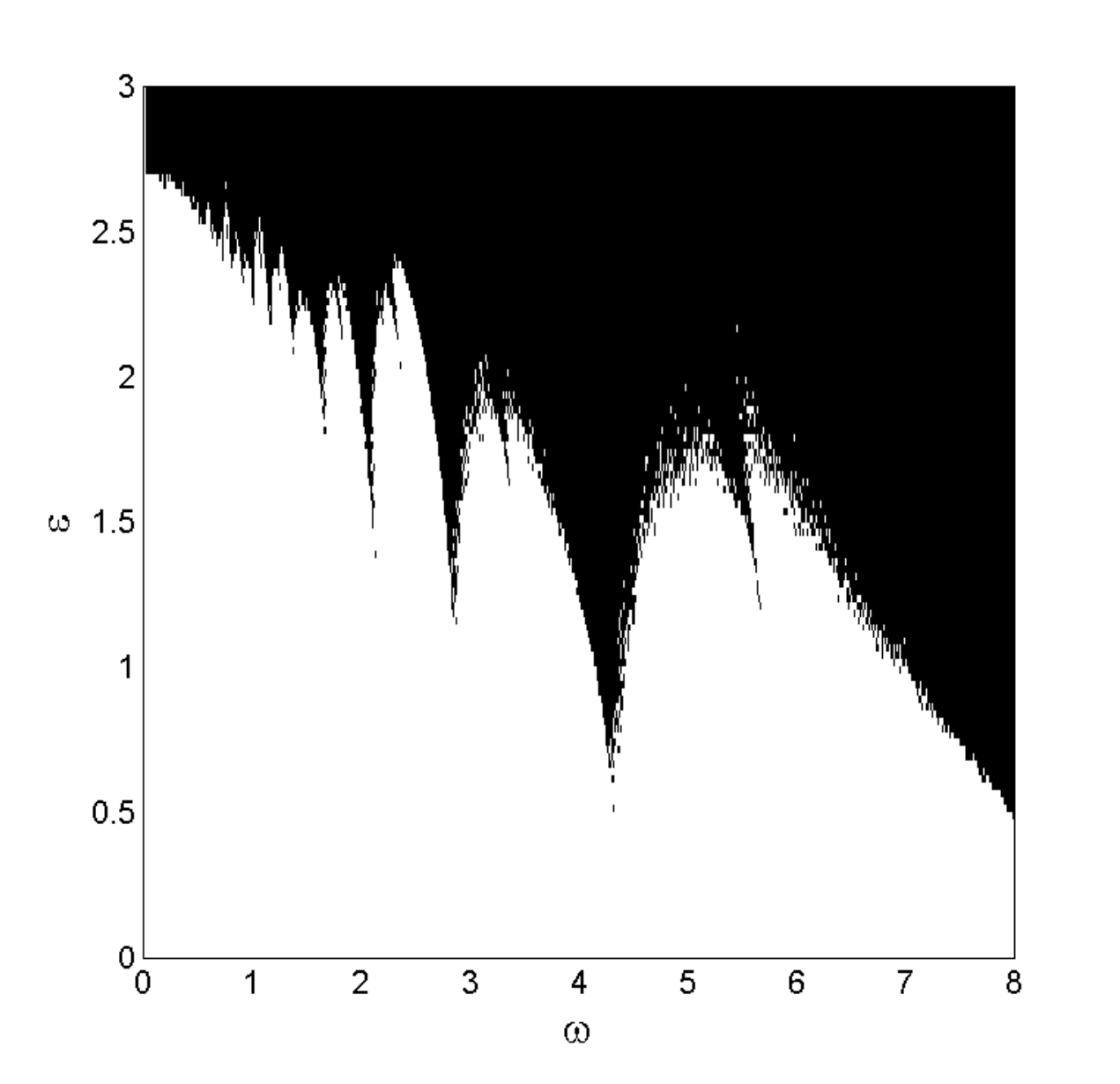}
\label{VaStabilityDiagram_SinCos_Wxy0p18N6p2}}
\subfigure[]{\includegraphics[width=3.4in]{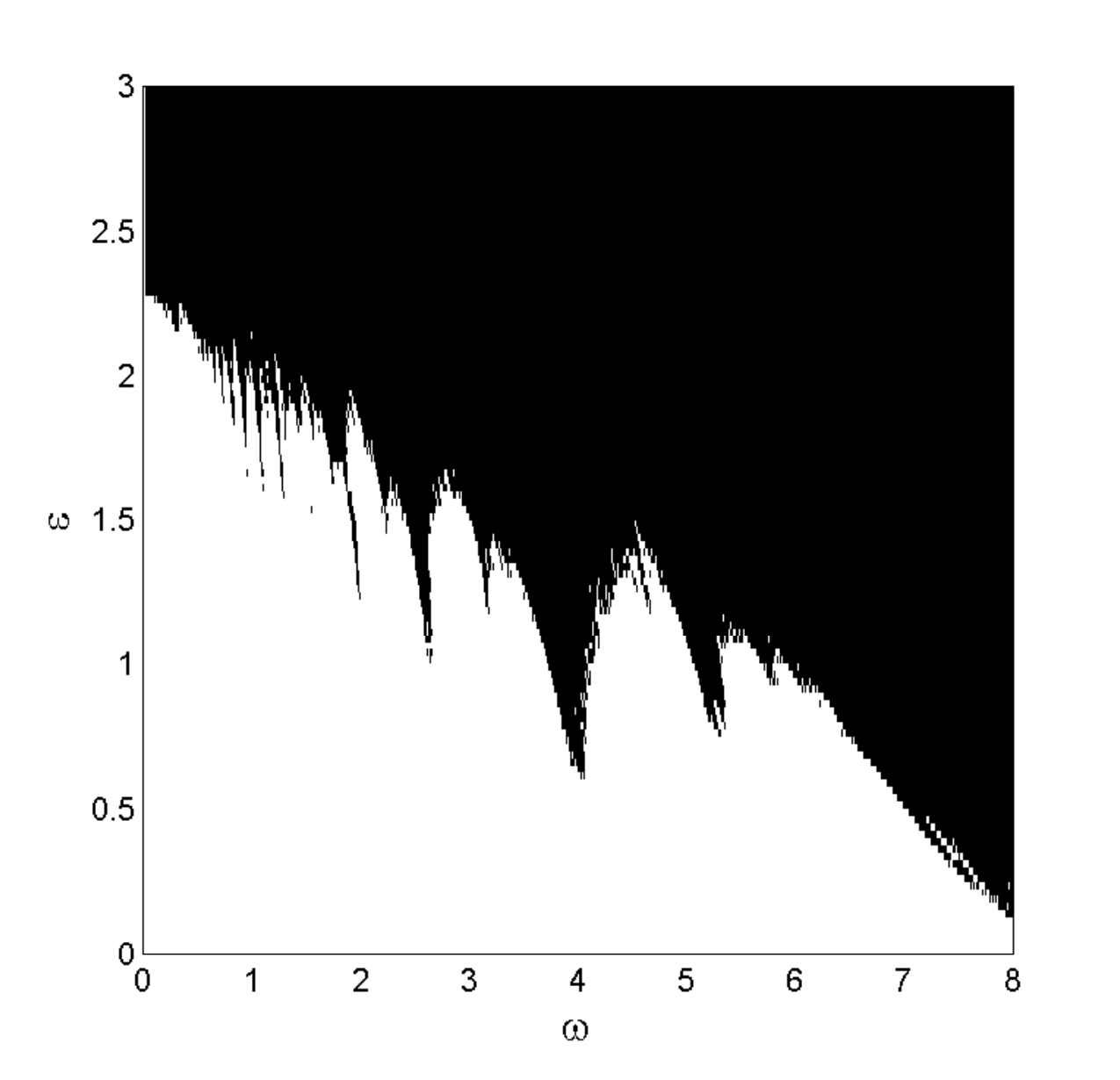}}
\label{VaStabilityDiagram_SinCos_Wxy0p28N5p29}}
\noindent\makebox[\textwidth]{
\subfigure[]{\includegraphics[width=3.4in]{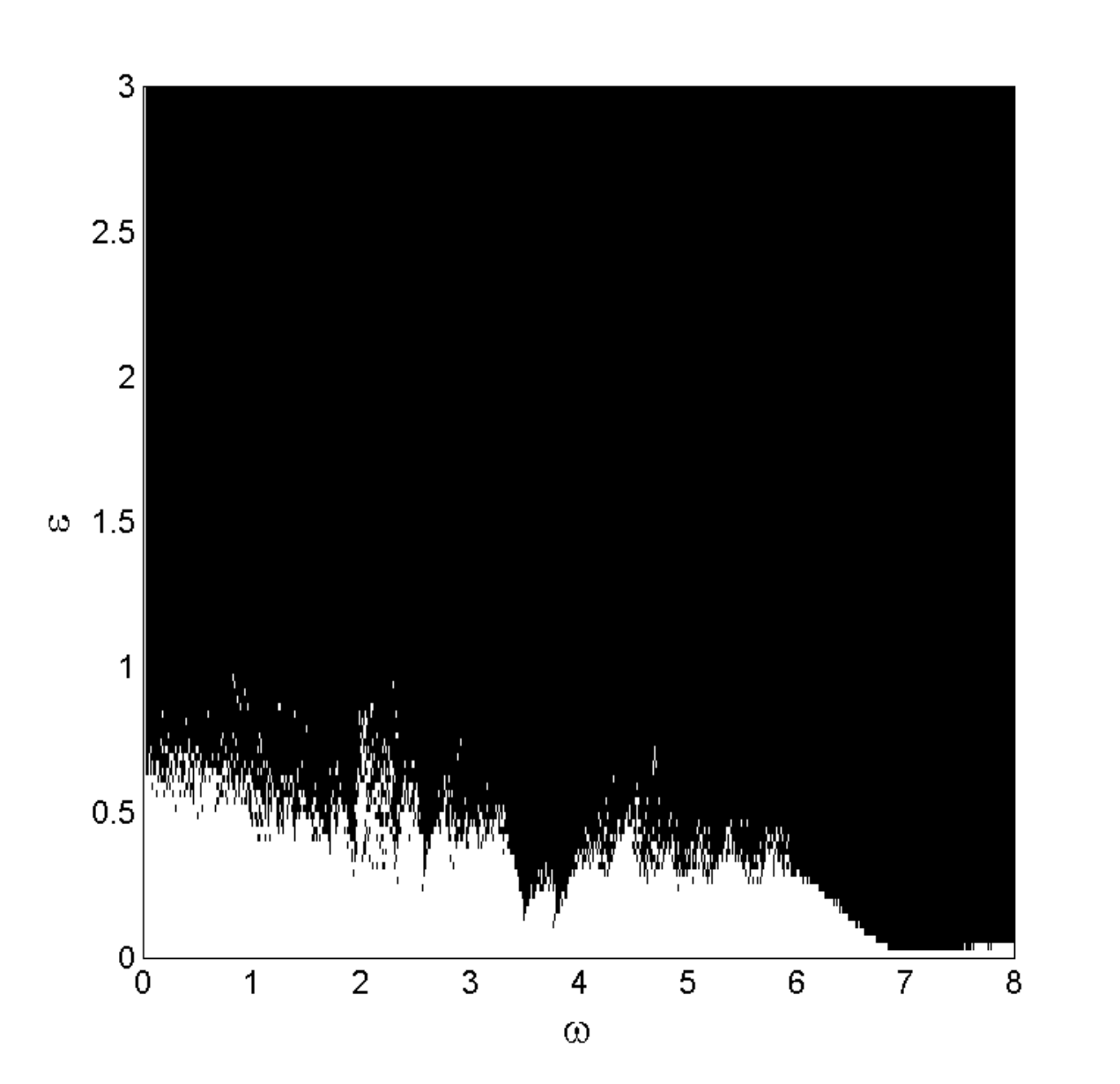}
\label{VaStabilityDiagram_SinCos_Wxy0p25N3p0}}
\subfigure[]{\includegraphics[width=3.4in]{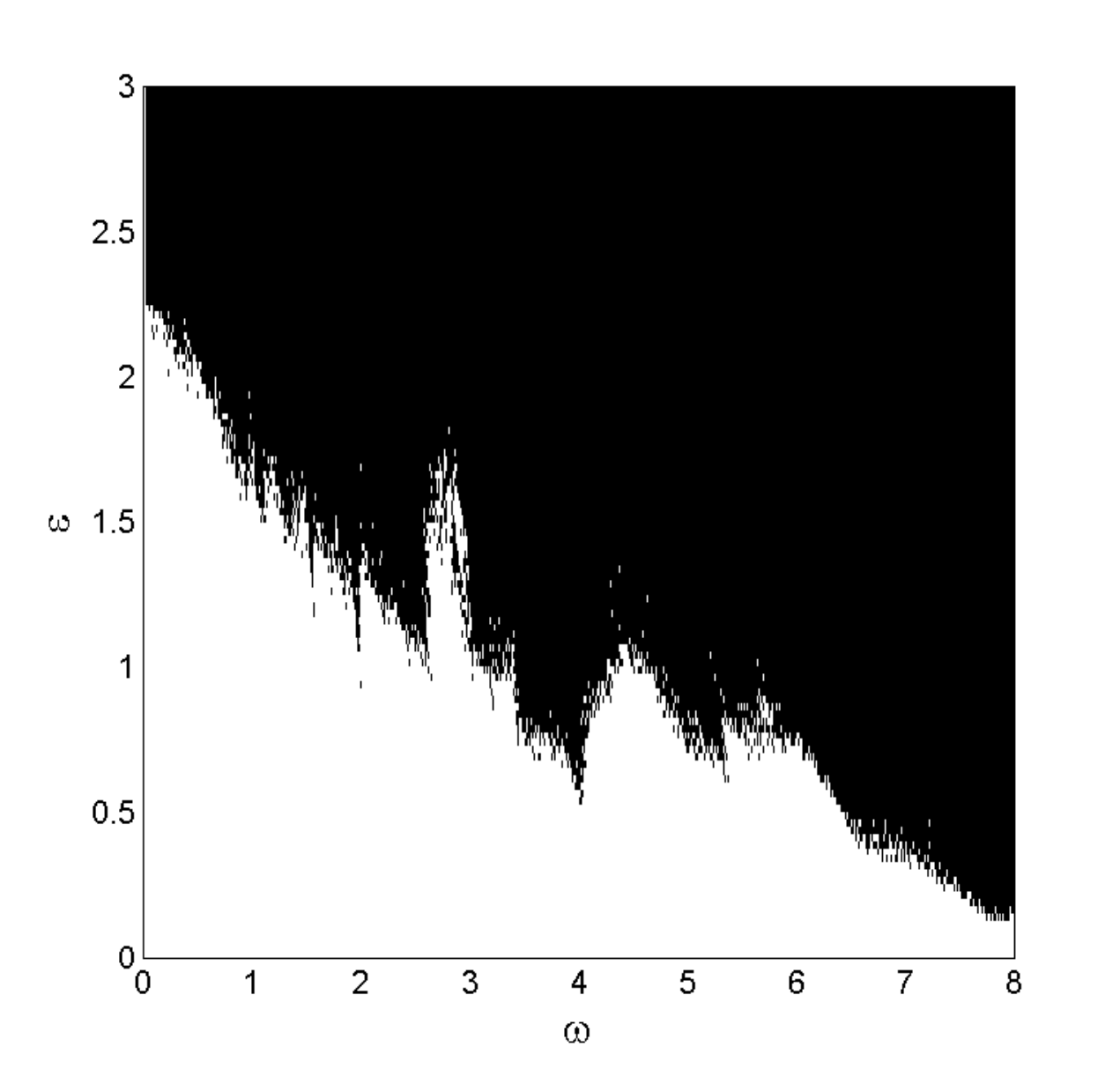}}
\label{VaStabilityDiagram_SinCos_Wxy0p22N5p29}}
\caption{The stability diagram predicted by the variational equations (%
\protect\ref{WidthVaEquations}) for the attractive nonlinearity and phase
shift $\protect\delta =\protect\pi /2$ between the modulations applied to
the $x$- and $y$-sublattices, in the plane of $\left( \protect\omega ,%
\protect\varepsilon \right) $. The initial conditions are given by Eqs. (%
\protect\ref{W0p18N6p2}), (\protect\ref{W0p28N5p29}), (\protect\ref{W0p25N3p0}%
) and (\protect\ref{W0p22N5p29}), for panels (a), (b), (c) and (d),
respectively.}
\label{VaStabilityDiagram_SinCos}
\end{figure}

When comparing the stable region in Fig.~\ref%
{VaStabilityDiagram_SinCos_Omega5Epsilon1} to its counterpart in the case of
the synchronous modulation, displayed in Fig.~\ref%
{VaStabilityDiagram_1PSin_Omega5Epsilon1}, significant differences are
observed. Specifically, the stability area is much smaller when the
time-modulation is phase-shifted by $\pi /2$. In fact, the latter area is
confined to large values of the norm and small widths. Several
representative examples if stability diagrams in the $(\omega ,\varepsilon )$
plane are presented in panels (a-d) of Fig.~\ref{VaStabilityDiagram_SinCos},
for parameter values
\begin{equation}
W_{x}=W_{y}=0.18,~N=6.2,  \label{W0p18N6p2}
\end{equation}%
\begin{equation}
W_{x}=W_{y}=0.28,~N=5.29,  \label{W0p28N5p29}
\end{equation}%
as well as for those given by Eqs. (\ref{W0p25N3p0}) and (\ref{W0p22N5p29}),
respectively. For parameter sets taken within the stable area of Fig.~\ref%
{VaStabilityDiagram_SinCos_Omega5Epsilon1}, such as the values given by Eq. (%
\ref{W0p18N6p2}), the stability pattern is optimized, exhibiting typical
peaks separated by well-defined regions of instability [Fig.~\ref%
{VaStabilityDiagram_SinCos}(a)]. Comparing the diagram in Fig. \ref%
{VaStabilityDiagram_SinCos}(a) to its counterpart in the case of the
synchronous modulation format [see Figs. \ref{VaStabilityDiagram_1PSin}(a,b)
and \ref{VaStabilityDiagram_1PSin_Exact}(a,b)], one can see that in the
present case, for low modulation frequencies, it is possible to achieve
stability beyond the critical threshold $\varepsilon _{\mathrm{cr}}$.
Furthermore, resonance frequencies, as found from careful examination of the
stability diagram in Fig. \ref{VaStabilityDiagram_SinCos_Wxy0p18N6p2}, are:

\begin{equation}
\omega _{0}^{\mathrm{(VA)}%
}=17.48,8.5,4.28,2.84,2.12,1.67,1.37,1.15,1.01,0.89,  \label{res-VA-sincos}
\end{equation}%
being rather similar to the ones obtained in the case of the
synchronous modulation, cf. Eq. (\ref{res-VA}). On the contrary, when
choosing points outside the stable domain, such as those corresponding to
Fig.~\ref{VaStabilityDiagram_SinCos}(c-d), stability peaks shrink and their
boundaries become fuzzy. Worth mentioning is the stability diagram produced
for the Gaussian ansatz with parameters given by Eq. (\ref{W0p22N5p29}),
which, as previously mentioned, corresponds to the stationary soliton with
parameters (\ref{V05Mum15}). This case, which refers to a point outside the stable area in Fig.~\ref{VaStabilityDiagram_SinCos_Omega5Epsilon1}, exhibits a rather disorderly stability pattern, where, in general, resonant frequencies cannot be immediately identified.
A similar scenario is observed for the same initial norm, but with a larger pulse's width, so that the corresponding point (\ref{W0p28N5p29}) is positioned near the edge of the stable domain in Fig.~\ref{VaStabilityDiagram_SinCos_Omega5Epsilon1}. In this case, the stability pattern, displayed in Fig.~\ref{VaStabilityDiagram_SinCos}(c), appears more structured, allowing the identification of several resonance frequencies,
\begin{equation}
\tilde{\omega}_{0}^{\mathrm{(VA)}}\cong 7.95,4.05,2.65,1.99,
\label{res-VA-sincos-W0p22N5p29}
\end{equation}%
(other frequencies are much harder to identify).

We have also collected results for modulation frequencies beyond the range presented in Fig.~\ref{VaStabilityDiagram_SinCos}, up to $\omega=20$. Focusing on the optimized setting in Eq. (\ref{W0p18N6p2}), a wide peak was found at $8.5<\omega <17.48$. We note that the height of this peak exceeds $\varepsilon_{cr}$, similar to the peaks found in any of the optimized stability diagrams produced by the synchronized modulation, as mentioned in Sec. \ref{sec:SelfAttractiveSymVA}.
For the parameters in both Eqs. (\ref{W0p22N5p29}) and (\ref{W0p28N5p29}), which might be more relevant for the comparison with the full numerical results (see below), this particular high-frequency stability peak attains height $\varepsilon_{max}\approx1$, significantly lower than the one seen in the optimized stability diagram.

\subsubsection{Numerical results}

\begin{figure}[tbp]
\centering
\subfigure[]{\includegraphics[width=3.4in]{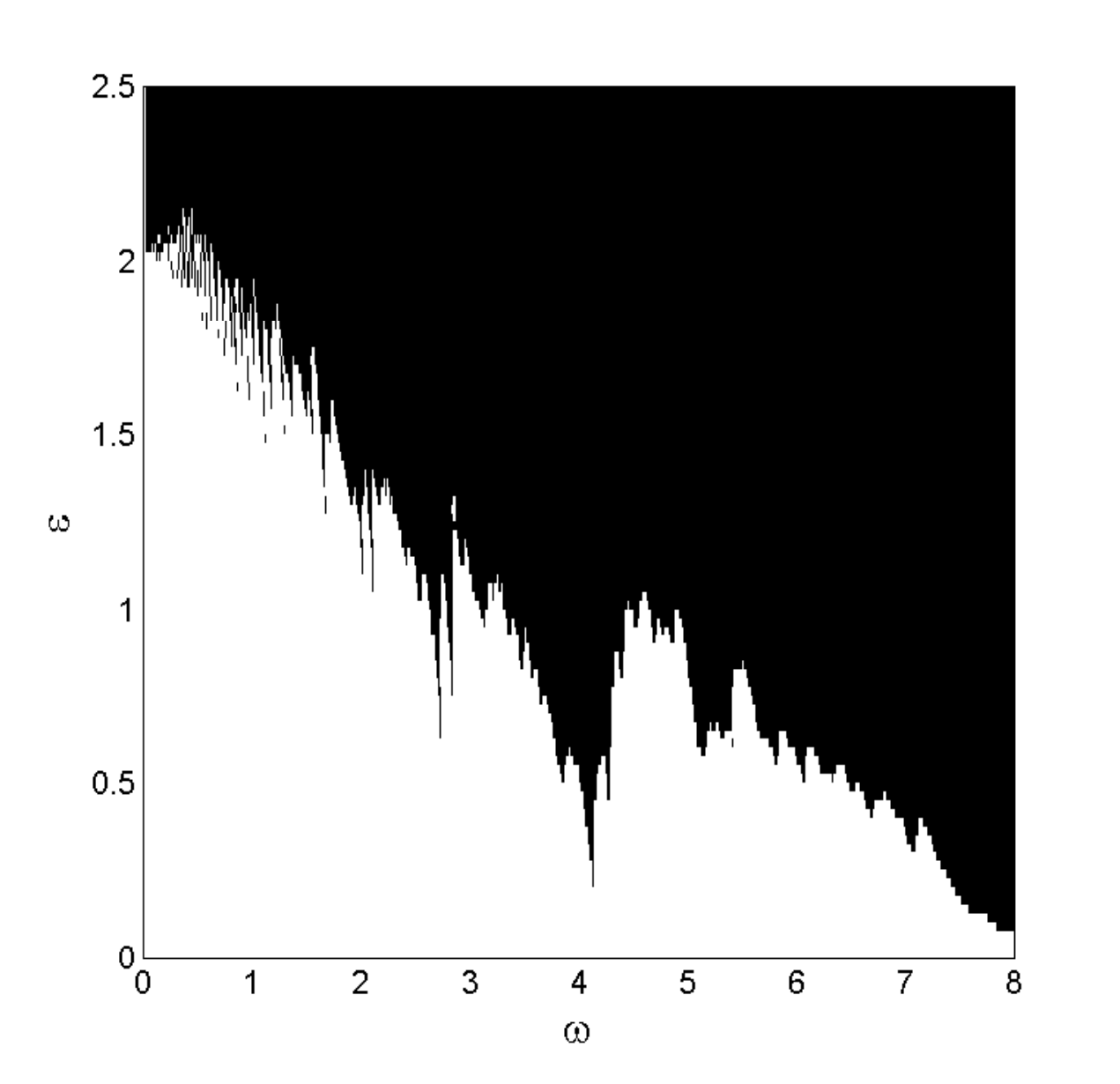}
\label{StabilityDiagram_SinCos_Wxy0p22N5p29}}
\caption{The same as in Fig.~\protect\ref{StabilityDiagram_1PSin}(b), but
with phase shift of $\protect\delta =\protect\pi /2$ between modulations
applied to the $x$- and $y$-sublattices.}
\label{StabilityDiagram_SinCos}
\end{figure}

Direct simulations of GPE (\ref{2DGPE}) were performed in this case as well.
They started with a numerically exact soliton solution for the static OL,
one of those used in Sec. \ref{sec:SelfAttractiveSymmetricModulation} for
parameters given by Eq. (\ref{V05Mum15}). The results are presented in Fig.~%
\ref{StabilityDiagram_SinCos} -- as before, in the $\left( \omega
,\varepsilon \right) $ plane.

Comparison between Fig.~\ref{StabilityDiagram_SinCos} and its counterpart
for the synchronously modulated 2D lattice displayed in Fig.~\ref%
{StabilityDiagram_1PSin_Wxy0p22N5p29} reveals some notable differences.
First, the $\pi /2$ phase shift causes a partial closure of the instability
tongues, and leads to fusion of originally separated stability peaks.
Further, at low modulation frequencies, stability may be observed for values of $\varepsilon$ beyond the critical threshold, $\varepsilon _{\mathrm{cr}}$, though these particular stability peaks are exceptionally narrow, as seen in Fig.~\ref{StabilityDiagram_SinCos}.
As mentioned in Sec. \ref{sec:SelfAttractiveSymmetricModulation}, the stability is not possible at $\varepsilon \geq \varepsilon _{\mathrm{cr}}$ for the synchronously modulated 2D\ lattice.

When comparing the numerically generated diagram displayed in Fig.~\ref{StabilityDiagram_SinCos} with the VA-predicted ones presented in Sec. \ref{sec:NonsynHalfPhiVA}, relatively close resemblance is found with the one seen in Fig.~\ref{VaStabilityDiagram_SinCos}(d), for parameter values (\ref{W0p22N5p29}) (as mentioned above, parameters (\ref{W0p22N5p29}) refer to the Gaussian function equivalent to the numerical solution produced by
the simulations with initial conditions (\ref{V05Mum15})), as well as with the stability diagram produced for a slightly larger initial width, using parameters (\ref{W0p28N5p29}), as seen in Fig.~\ref{VaStabilityDiagram_SinCos}(b). In fact, the stability pattern in Fig.~\ref{StabilityDiagram_SinCos} appears as an intermediate one, between two approximate patterns displayed in Figs. ~\ref{VaStabilityDiagram_SinCos}(d) and ~\ref{VaStabilityDiagram_SinCos}(b). In addition, four resonance frequencies, similar to ones given by Eq. (\ref
{res-VA-sincos-W0p22N5p29}), can be extracted from Fig.~\ref{StabilityDiagram_SinCos}: $\tilde{\omega}_{0}^{\mathrm{(num)}}\cong 8.21,4.13,2.73,2.02$.

Direct numerical simulations of GPE (\ref{2DGPE}) were also performed for high modulation frequencies (not shown in Fig.~\ref{StabilityDiagram_SinCos}), revealing a wider stability peak, roughly in the range of $8.21<\omega <15.90$. Comparing this outcome with the one obtained for the synchronized setting (Sec. \ref{sec:SelfAttractiveSymNum}), we note that this high-frequency peak is significantly shrunk, extending no higher than to $\varepsilon_{max}=0.68$ (while the height of the equivalent peak in the synchronous configuration nearly reaches $\varepsilon_{cr}$). Actually, the decrease in the height of this stability peak was predicted by the VA, see Sec. \ref{sec:NonsynHalfPhiVA}.

The investigation of this configuration can be extended further. In
particular, we here do not examine the possibility of improving the
stability results by considering an initial solution with a larger norm,
i.e., for solitons taken deeper in the semi-infinite bandgap, which is
possible according to the prediction of the VA analysis. We do not attempt
either to use initial conditions other than the numerically exact stationary
soliton produced by solving the GPE (\ref{2DGPE}) with the static OL [Eq. (%
\ref{Basic2DExternalPotential})]. A different approach may be to modify the
input, using Gaussians with larger norms and smaller widths. In the latter
case, the stability criterion adopted in the present work ($99\%$
conservation of the initial norm) may need to be modified, as some excess
norm is expected to be shed off by such a soliton in the course of its
evolution.

\subsection{The phase shift of $\protect\delta =\protect\pi $ between the
sublattices}

\subsubsection{Variational results}
\label{sec:NonsynPhiVA}
We have examined the modulation pattern with the phase shift of $\pi $ in
Eq. (\ref{General2DExternalPotential}), following the scenario elaborated
above for the cases of $\delta =0$ (synchronous modulation of the 2D
lattice) and $\delta =\pi /2$. The stability chart, displayed in Fig.~\ref%
{VaStabilityDiagram_SinMSin_Omega5Epsilon1}, was plotted for the same
reference point (\ref{RefPointOmega5Epsilon1}) as employed above, using the
variational equations (\ref{WidthVaEquations}). While the respective
stability chart for $\delta =\pi /2$, shown in Fig.~\ref%
{VaStabilityDiagram_SinCos_Omega5Epsilon1}, displays a significant shrinkage
of the stability area, in the present case the stability chart, displayed~in
Fig. \ref{VaStabilityDiagram_SinMSin_Omega5Epsilon1}, is relatively close to
the one constructed for the synchronous time presented above in Fig.~\ref%
{VaStabilityDiagram_1PSin_Omega5Epsilon1}.

\begin{figure}[tbp]
\centering
\subfigure[]{\includegraphics[width=3.4in]{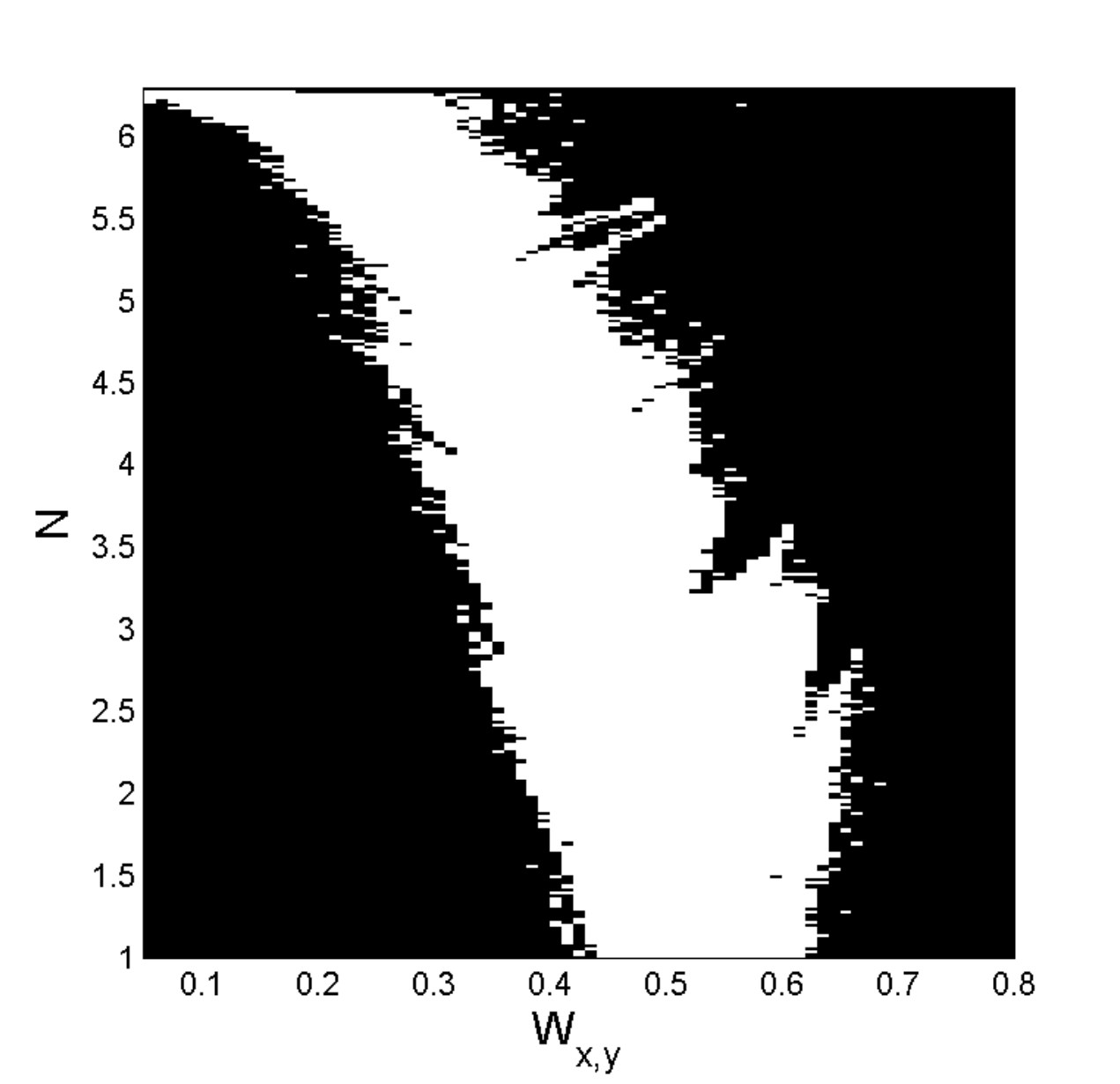}
\label{VaStabilityDiagram_SinMSin_Omega5Epsilon1}}
\caption{The stability chart\ in the plane of $\left( W_{x}=W_{y},N\right) $,
similar to the one in Fig.~\protect\ref%
{VaStabilityDiagram_SinCos_Omega5Epsilon1}, but for phase shift $\protect%
\delta =\protect\pi $ between the modulations of the $x$ and $y$ sublattices.
}
\label{VaStabilityDiagram_SinMSin_Omega5Epsilon1}
\end{figure}

\begin{figure}[tbp]
\centering
\noindent\makebox[\textwidth]{
\subfigure[]{\includegraphics[width=3.4in]{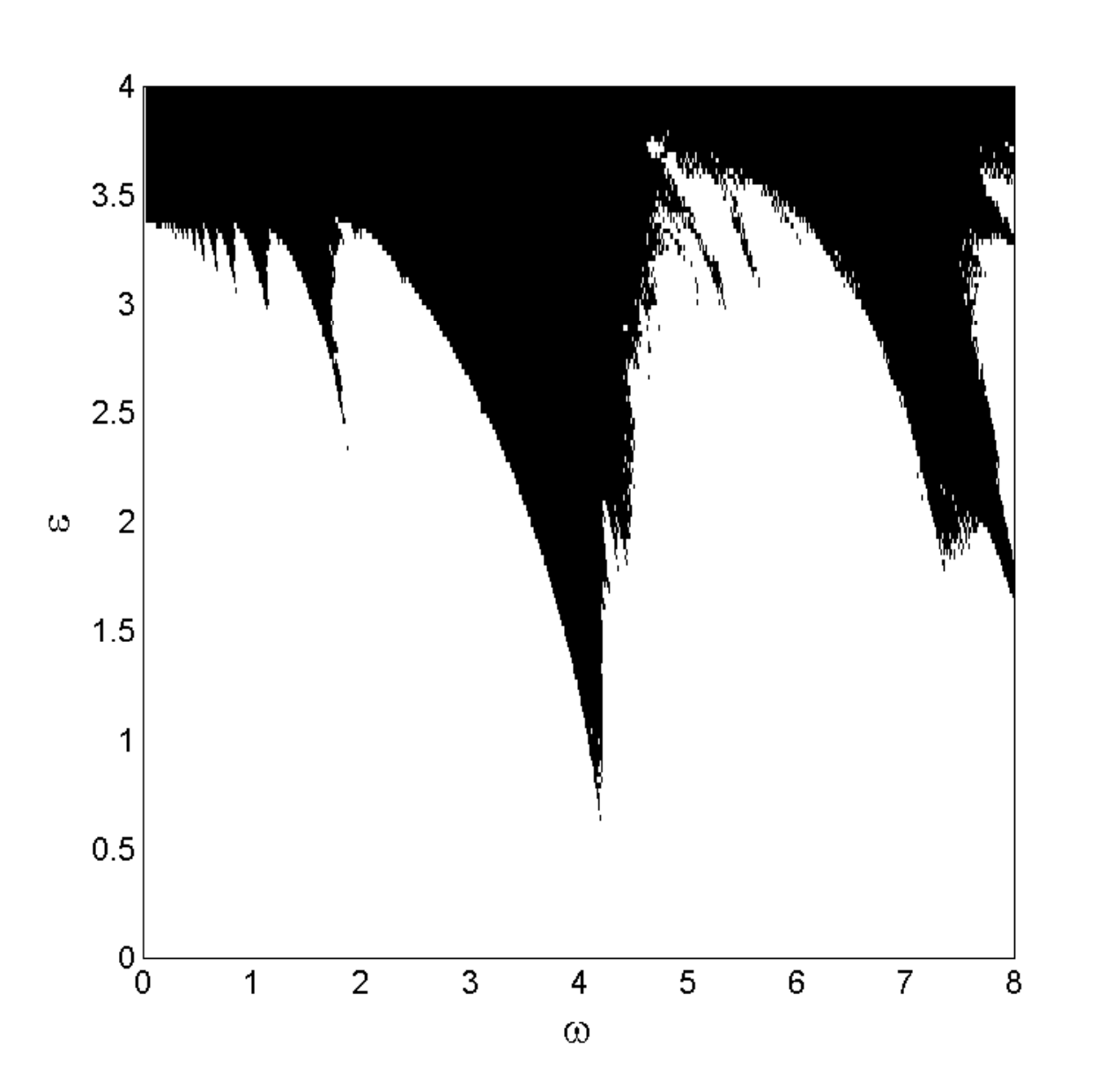}
\label{VaStabilityDiagram_SinMSin_Wxy0p25N5p8}}
\subfigure[]{\includegraphics[width=3.4in]{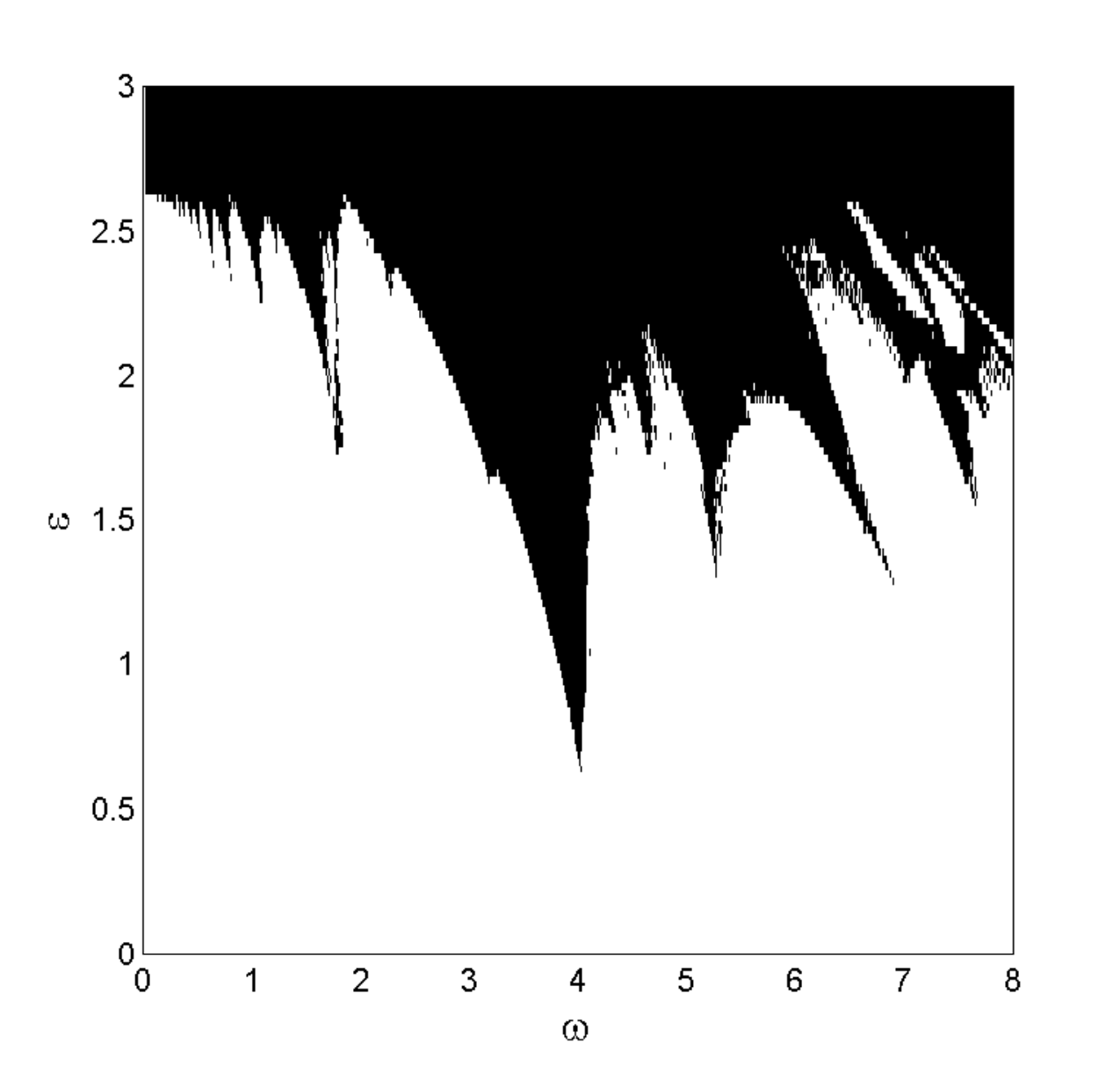}}
\label{VaStabilityDiagram_SinMSin_Wxy0p3N5p29}}
\noindent\makebox[\textwidth]{
\subfigure[]{\includegraphics[width=3.4in]{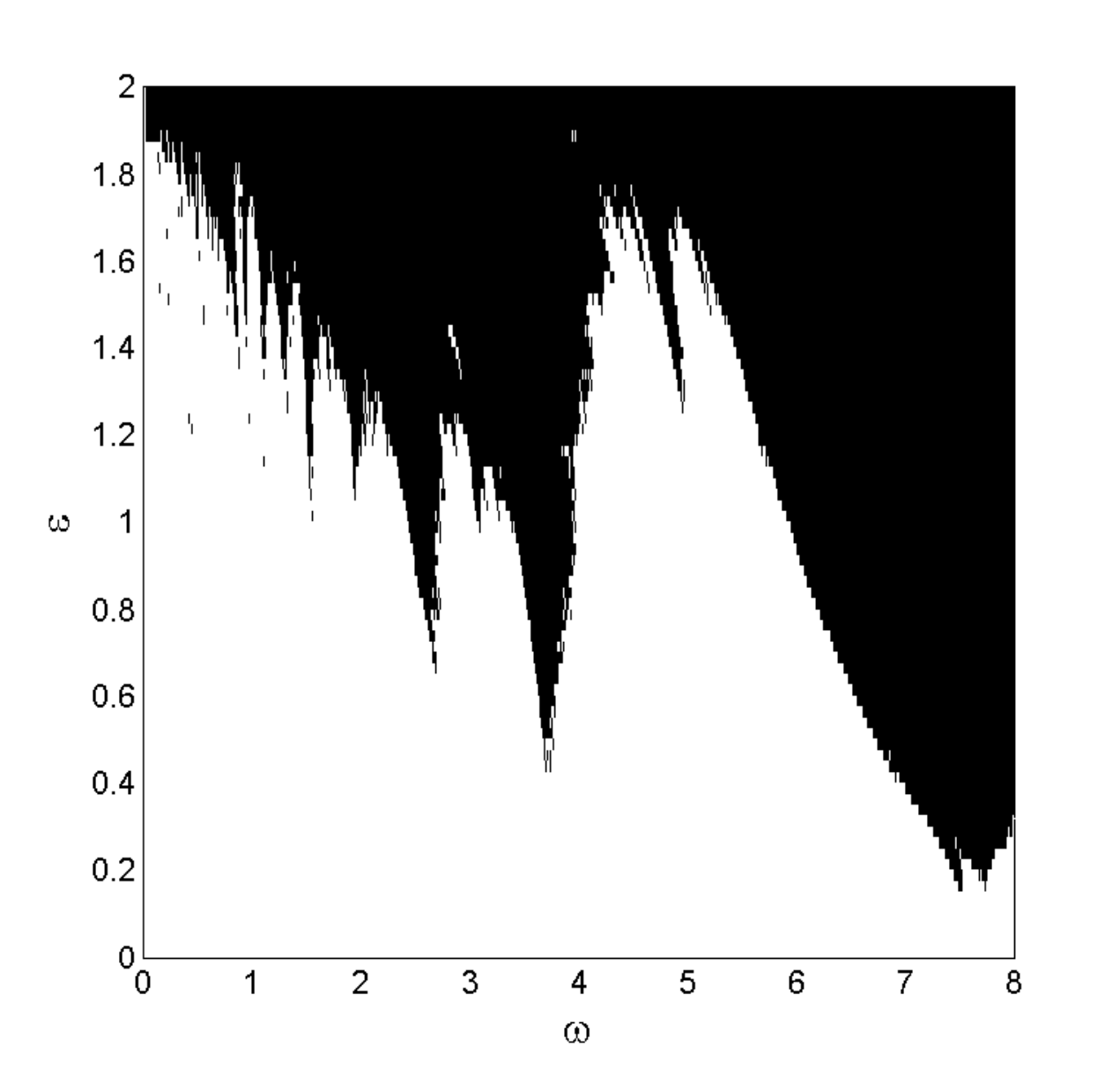}
\label{VaStabilityDiagram_SinMSin_Wxy0p5N2p0}}
\subfigure[]{\includegraphics[width=3.4in]{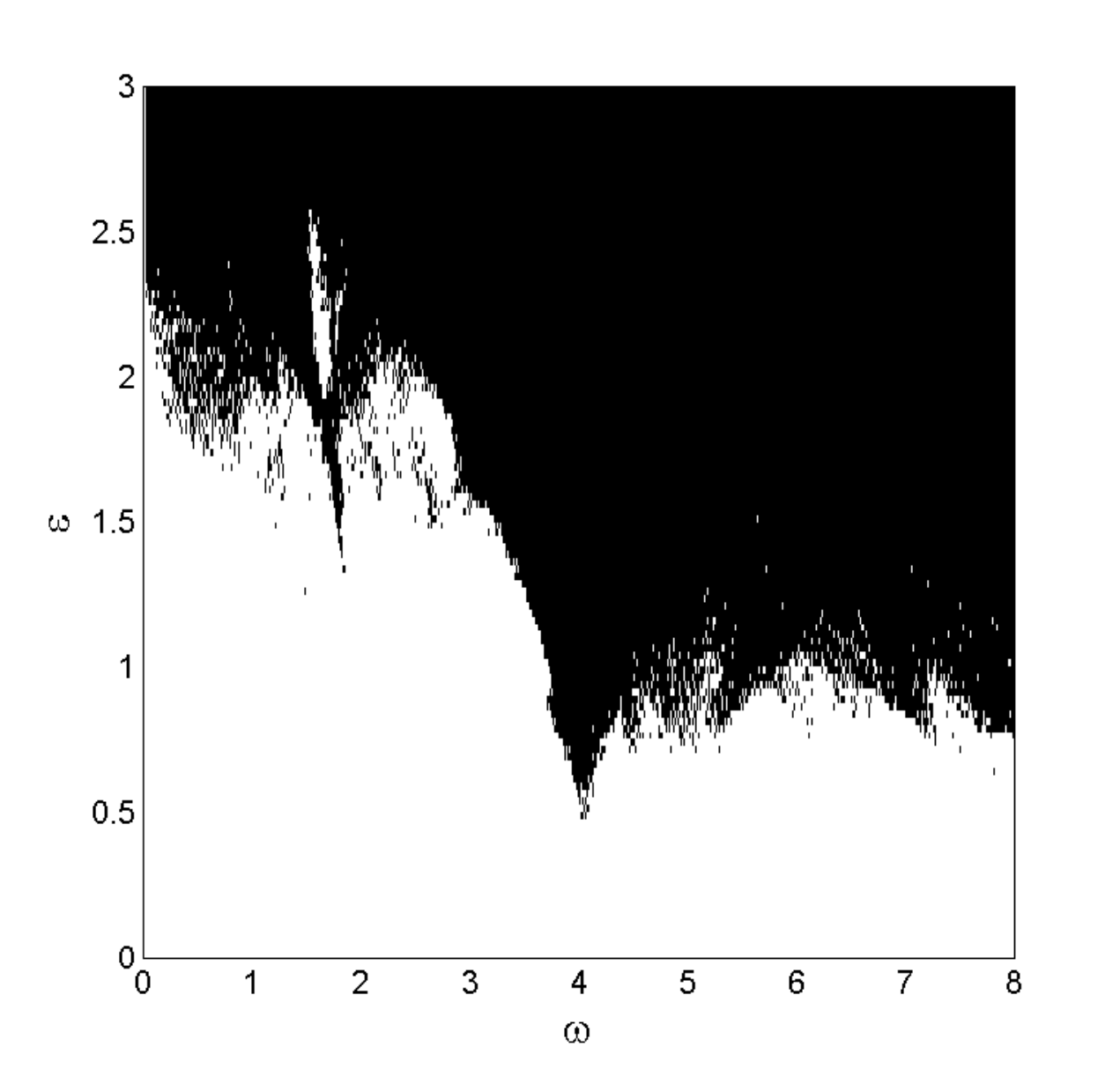}}
\label{VaStabilityDiagram_SinMSin_Wxy0p22N5p29}}
\caption{Examples of stability diagrams, predicted by the variational
equations (\protect\ref{WidthVaEquations}), with the attractive nonlinearity
and phase shift $\protect\delta =\protect\pi $ between modulations of the $x$%
- and $y$-sublattices, in the plane of $\left( \protect\omega ,\protect%
\varepsilon \right) $. Panels (a), (b), (c) and (d) correspond to the
initial conditions defined by Eqs. (\protect\ref{W0p25N5p8}), (\protect\ref{W0p3N5p29}), (\protect\ref{W0p5N2p0}), and (\protect\ref{W0p22N5p29}),
respectively.}
\label{VaStabilityDiagram_SinMSin}
\end{figure}

Several examples of stability diagrams in the plane of the modulation
parameters, $\left( \varepsilon ,\omega \right) $, for several
representative points in the chart~displayed in Fig. \ref%
{VaStabilityDiagram_SinMSin_Omega5Epsilon1},
\begin{equation}
W_{x}=W_{y}=0.25,N=5.8,  \label{W0p25N5p8}
\end{equation}%
\begin{equation}
W_{x}=W_{y}=0.3,N=5.29,  \label{W0p3N5p29}
\end{equation}%
\begin{equation}
W_{x}=W_{y}=0.5,N=2.0,  \label{W0p5N2p0}
\end{equation}%
and the parameter set given by Eq. (\ref{W0p22N5p29}%
), are plotted in Fig.~\ref{VaStabilityDiagram_SinMSin}(a-d).

As expected, for parameters positioned in the center of the stability area
in Fig.~\ref{VaStabilityDiagram_SinCos_Omega5Epsilon1}, with a sufficiently large initial norm, such as one given by Eq. (\ref{W0p25N5p8}), the respective stability diagrams are optimized ones, featuring large and distinctive stability regions.
In the particular case of parameters (\ref{W0p25N5p8}), the stability region covers almost entirely the physically significant domain, $0\leq \varepsilon \leq 2$, for the explored modulation-frequency range, $\omega <20$, with the exception of instability tongues around the first three lower-order resonant frequencies [see Eq. (\ref{res-VA-sinMsin})], two of which can be seen in Fig.~\ref{VaStabilityDiagram_SinMSin}(a).
The stability pattern plotted in Fig.~\ref{VaStabilityDiagram_SinMSin}(a)
actually represents scenarios for which the stability domains are broadest,
in comparison with all the other configurations examined in this study.

The stability investigation was also performed for initial parameters (\ref{W0p3N5p29}), which correspond to a Gaussian pulse with the norm equal to that of the exact solution used as the input in the direct numerical simulations (detailed in the section below), but with a larger width, still remaining at the center of the stability area in Fig.~\ref{VaStabilityDiagram_SinMSin_Omega5Epsilon1}. Here, as shown in Fig.~\ref{VaStabilityDiagram_SinMSin}(b), a relatively clear stability profile is maintained at low modulation frequencies, $\omega\lesssim4$, while at higher frequencies the stability pattern becomes disorganized. In more detail, as partially shown in Fig.~\ref{VaStabilityDiagram_SinMSin}(b), an ``untidy" peak originates at resonant frequency $\omega=4.03$, and is extended further than usual, up to frequency $\omega\cong13$.
For even smaller values of the norm, the stability pattern regains its well-structured form, even at high frequencies considered in Fig~\ref{VaStabilityDiagram_SinMSin}, though the stability peaks are significantly reduced. Typical example of such a case is presented in Fig.~\ref{VaStabilityDiagram_SinMSin}(c), for parameters (\ref{W0p5N2p0}).

Near edges of the stability area in Fig. %
\ref{VaStabilityDiagram_SinMSin_Omega5Epsilon1}, like those corresponding to
the set of initial values given by Eq. (\ref{W0p22N5p29}), the stability
regions shrink and merge, see Fig.~\ref{VaStabilityDiagram_SinMSin}(d).
Outside the stability area in Fig.~\ref%
{VaStabilityDiagram_SinMSin_Omega5Epsilon1}, the stability regions
experience a substantial shrinkage (for parameters defined as per Eq. (\ref{W0p25N3p0}), the stability diagram is very similar to one for the $\pi/2$ phase shift, displayed in Fig.~\ref{VaStabilityDiagram_SinCos}(c)).

The resonance frequencies, as obtained from the optimized VA-based
diagram in Fig.~\ref{VaStabilityDiagram_SinMSin}(a), are:
\begin{equation}
\omega _{0}^{\mathrm{(VA)}}=18.15,8.43,4.2,1.88,1.14,0.85,0.67,0.55,0.47,0.41.
\label{res-VA-sinMsin}
\end{equation}%
When comparing the
frequencies from Eq. (\ref{res-VA-sinMsin}) to the variationally predicted
ones for the synchronous time-modulation scheme, see Eq. (\ref{res-VA}) (and, actually, also to the numerically obtained resonant frequencies specified in the following section), large differences are seen for the low resonant frequencies. These results
may be explained by noting that the stability peaks (and instability
tongues), observed in Fig.~\ref{VaStabilityDiagram_SinMSin}(a) -- in
particular, the ones created at low modulation frequencies -- are not
entirely vertical, but tilted at a certain angle. This observation suggests that the resonant
frequencies are not constant, but depend on the modulation frequency $%
\varepsilon $, although the dependence becomes conspicuous only at
sufficiently large values of $\varepsilon $. Such a property was not seen in
any of the configurations examined above. Here, the instability tongues that
correspond to low resonant frequencies appear only at $\varepsilon >2.5$,
well above values considered in the stability patterns that we examined above.
It should also be mentioned that the low resonant frequencies in Fig.~\ref{VaStabilityDiagram_SinMSin}(b) [for parameters (\ref{W0p3N5p29})], i.e., belonging to the structured part of the stability profile, are fairly close to ones in Eq. (\ref{res-VA-sinMsin}), for the case of large norm, with the difference of $<5\%$.

Similarly to the time-modulation profiles discussed above, the stability was also examined here for modulation frequencies higher than those presented in Fig.~\ref{VaStabilityDiagram_SinMSin}, {\it viz.}, at $\omega>8$. As implied by Eq. (\ref{res-VA-sinMsin}), for parameters (\ref{W0p25N5p8}) that represent the optimal stability scenario, a stability peak exists in the range of $8.43<\omega<18.15$, stretching beyond $\varepsilon=3.5$.
For parameters (\ref{W0p3N5p29}), which may be more appropriate for comparison with the exact numerical results, the high-frequency peak exists approximately at $13\lesssim\omega\lesssim22.5$, extending up to $\varepsilon=2.55$ and, similar to the previous one, it is also not as neat as some of the low-frequencies peaks.

\subsubsection{Numerical results}

\begin{figure}[tbp]
\centering
\subfigure[]{\includegraphics[width=3.4in]{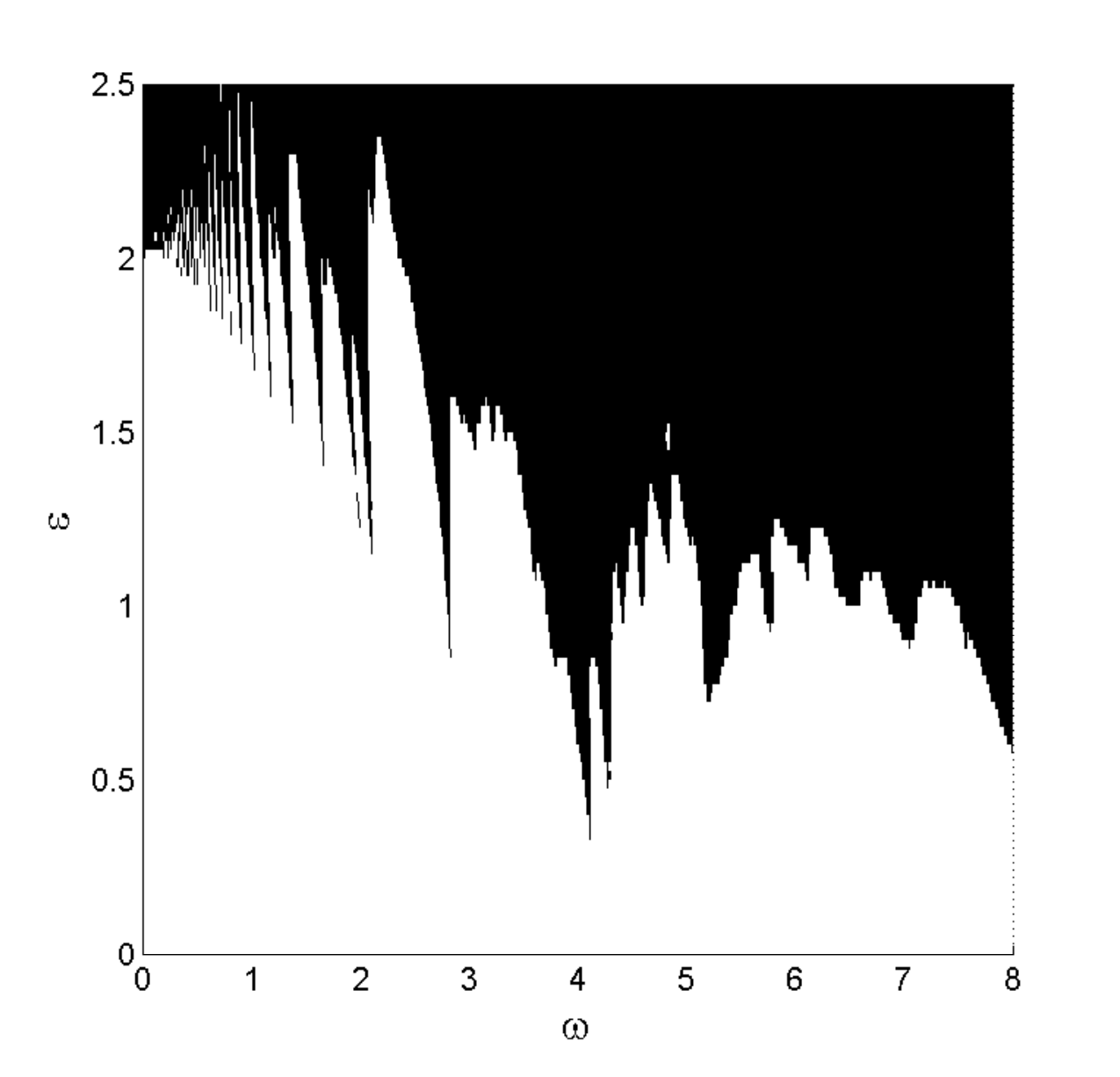}
\label{StabilityDiagram_SinMSin_Wxy0p22N5p29}}
\caption{The stability diagram produced by direct simulations of Eq. (%
\protect\ref{2DGPE}), similar to diagrams plotted in Figs. \protect\ref%
{StabilityDiagram_1PSin}(b) and \protect\ref{StabilityDiagram_SinCos}, but
with the phase shift of $\protect\delta =\protect\pi $ between modulations
of the $x$- and $y$-sublattices.}
\label{StabilityDiagram_SinMSin}
\end{figure}

Results of simulations of GPE (\ref{2DGPE}) for $\delta =\pi $ are displayed
in Fig.~\ref{StabilityDiagram_SinMSin}. As above, numerically exact
stationary solutions of Eq. (\ref{V05Mum15}) were used as initial condition
for the simulations. Comparison of the stability pattern in Fig.~\ref%
{StabilityDiagram_SinMSin} with one plotted for the synchronously modulated
2D sublattice in Fig.~\ref{StabilityDiagram_1PSin_Wxy0p22N5p29} demonstrates
an increase of $\approx 20\%$ in the height of the stability peaks within the region $0<\omega<8$.
The resonant frequencies, revealed by the inspection of
the numerical findings (as usual, the first ten),
\begin{equation}
\omega _{0}^{\mathrm{(num)}}=15.80,8.54,4.11,2.82,2.1,1.66,1.37,1.17,1.02,0.9,
\label{res-num-sinMsin}
\end{equation}%
are very close to those found for the synchronously modulated lattice in (%
\ref{res-num}), with differences $<2\%$.

In contrast with the time-modulation profiles examined above, in the present case it is more difficult to identify common characteristics between the stability pictures produced through direct simulations of the GPE and the
VA-predicted ones.
As mentioned above, the respective variational stability diagram in Fig.~\ref%
{VaStabilityDiagram_SinMSin}(c), with parameters corresponding to Eq. (\ref%
{W0p22N5p29}), shows a relatively disordered stability pattern, quite
different from the one in Fig.~\ref{StabilityDiagram_SinMSin}.
As concerns the optimized VA diagrams demonstrated in Fig.~\ref{VaStabilityDiagram_SinMSin}(a) (and, to a certain extent, the diagram in Fig.~\ref{VaStabilityDiagram_SinMSin}(b) as well), the anticipated form of tilted stability peaks is not observed in the numerically generated pictures. For that reason, the resonant frequencies in Eq. (\ref{res-num-sinMsin}) are quite different from their variationally predicted counterparts in Eq. (\ref{res-VA-sinMsin}).
Nevertheless, the predicted growth of the stability peaks, revealed by comparing the optimized VA diagrams, plotted in Fig.~\ref{VaStabilityDiagram_SinMSin}(a), with their variational counterparts for the synchronously modulated 2D lattice, see Fig.~\ref{VaStabilityDiagram_1PSin}(a,b), turns out to be
generally correct in the low-frequency domain ($\omega<8$).

Analysis of the stability at high frequencies, up to $\omega=20$, has revealed a conspicuously reduced stability region at $8.54<\omega<15.8$, which, at its highest, approaches $\varepsilon_{max}=1.04$.
This shallow stability peak is somewhat higher than the similar one observed in the case of the $\pi/2$ phase shift between the sublattices. On the other hand, when comparing to the results obtained with the synchronously modulated sublattices, a considerable shrinkage still happens for the $\pi$ phase shift.
This outcome is not predicted by the VA, as the corresponding high-frequency stability peak is too steep for the use of the optimization provided by Eq. (\ref{W0p25N5p8}) [and also Eq. (\ref{W0p3N5p29})], as detailed in Sec. Sec.~\ref{sec:NonsynPhiVA}.

\section{Vortices in synchronously modulated 2D lattices}

\label{sec:VorticesStyability}

\begin{figure}[tbp]
\centering
\noindent\makebox[\textwidth]{
\subfigure[]{\includegraphics[width=3.0in]{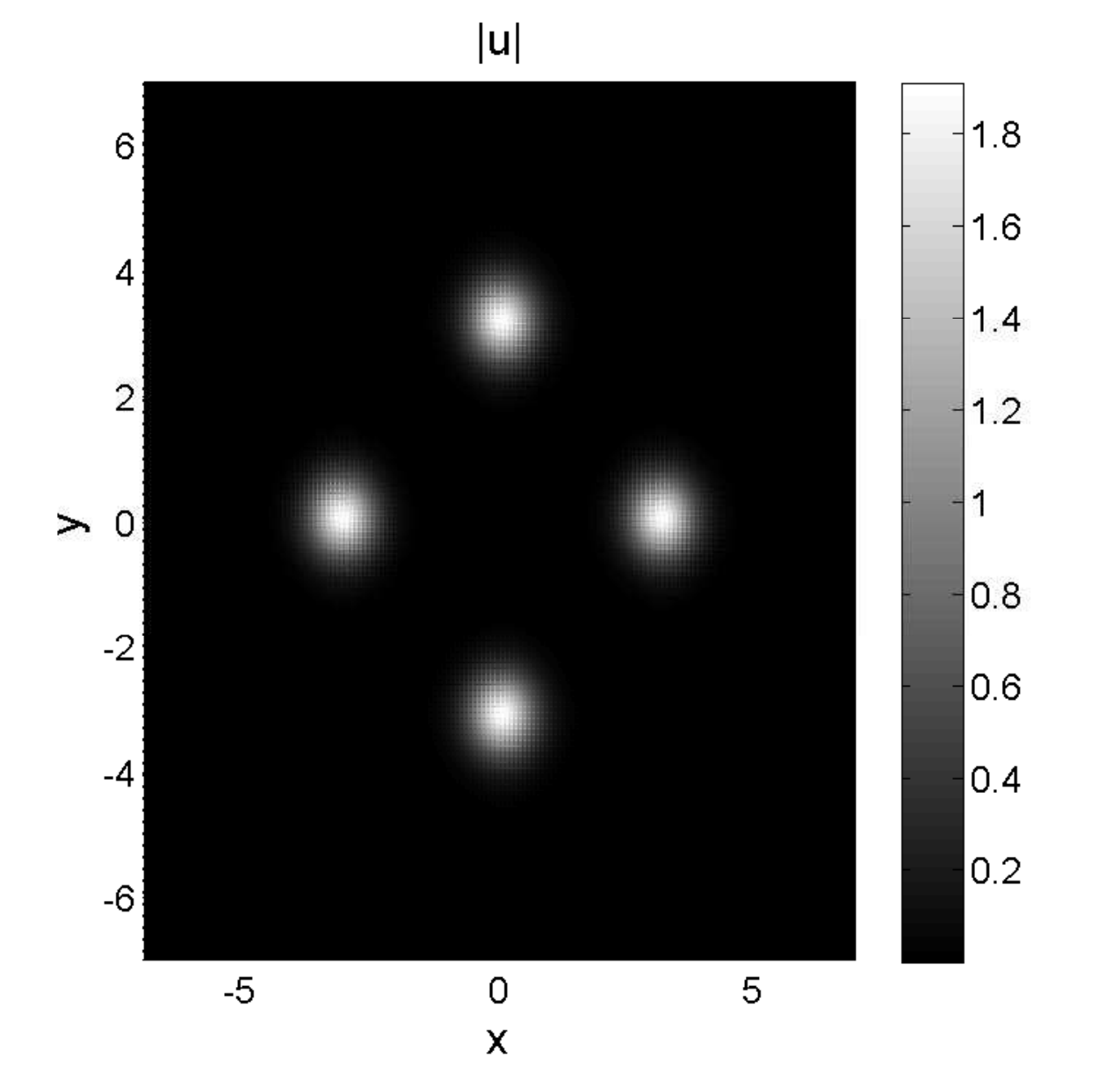}
\label{RhombicAmpProfile}}
\subfigure[]{\includegraphics[width=3.0in]{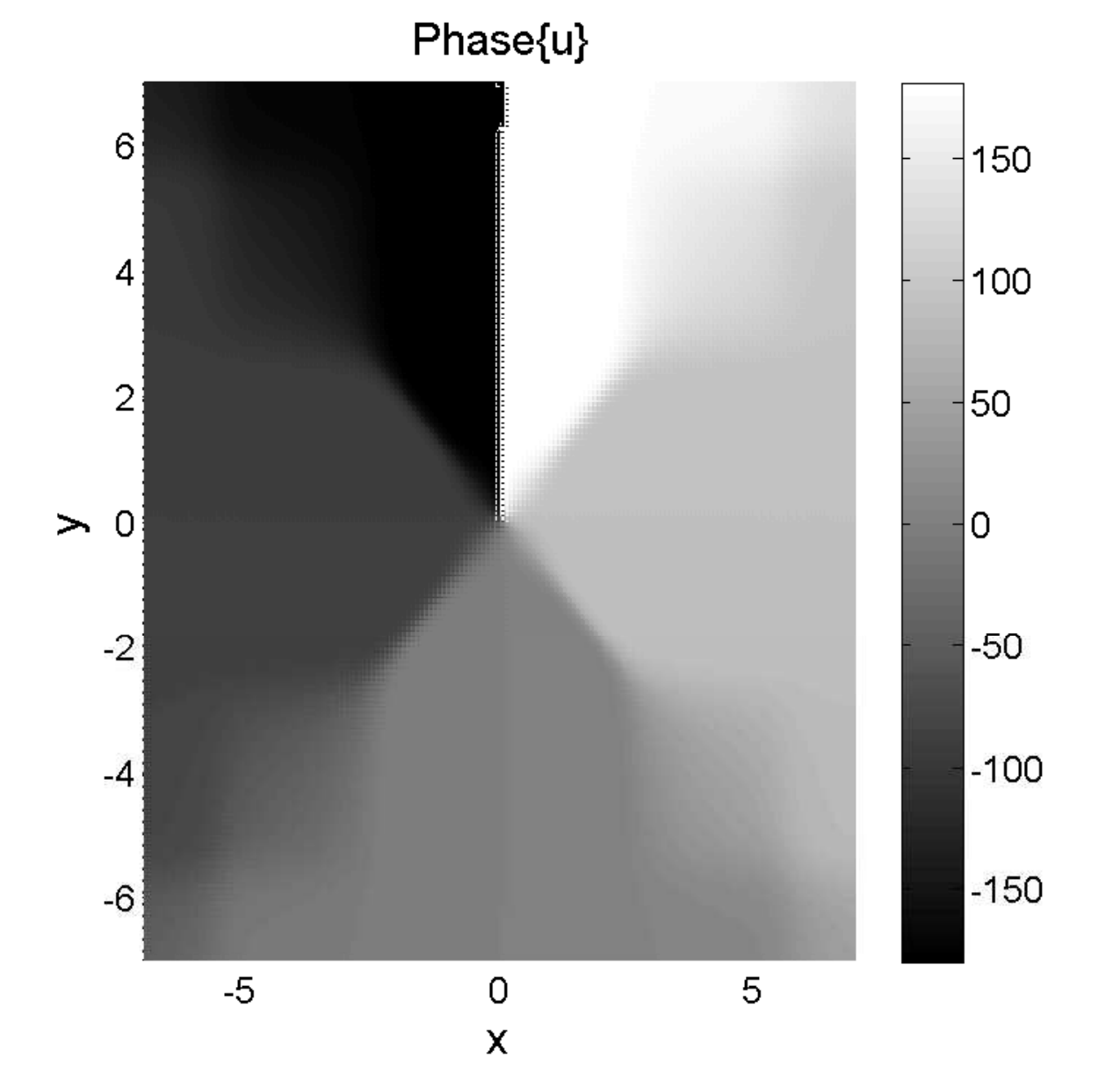}}
\label{RhombicPhaseProfile}}
\noindent\makebox[\textwidth]{
\subfigure[]{\includegraphics[width=3.0in]{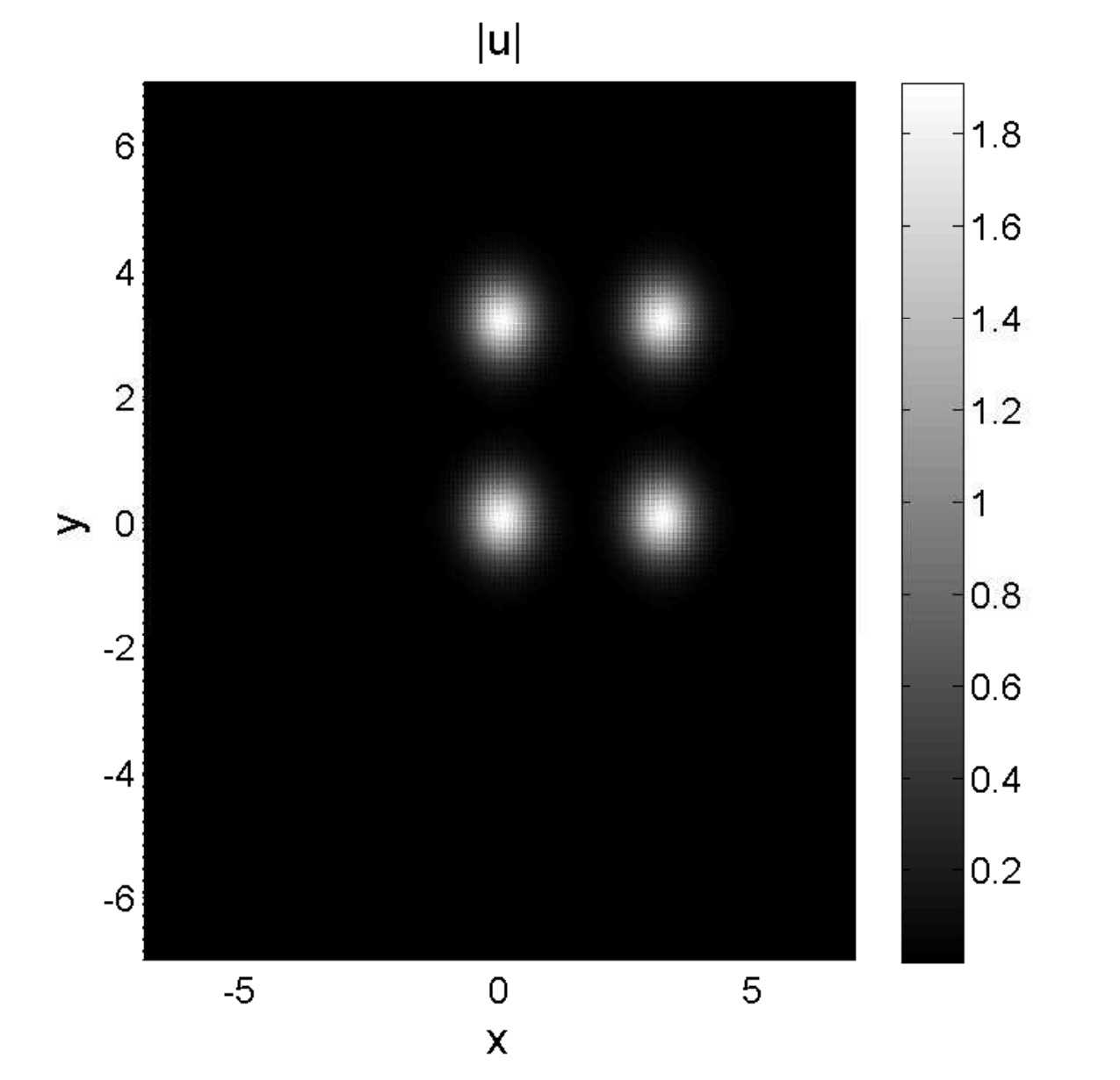}
\label{SquareAmpProfile}}
\subfigure[]{\includegraphics[width=3.0in]{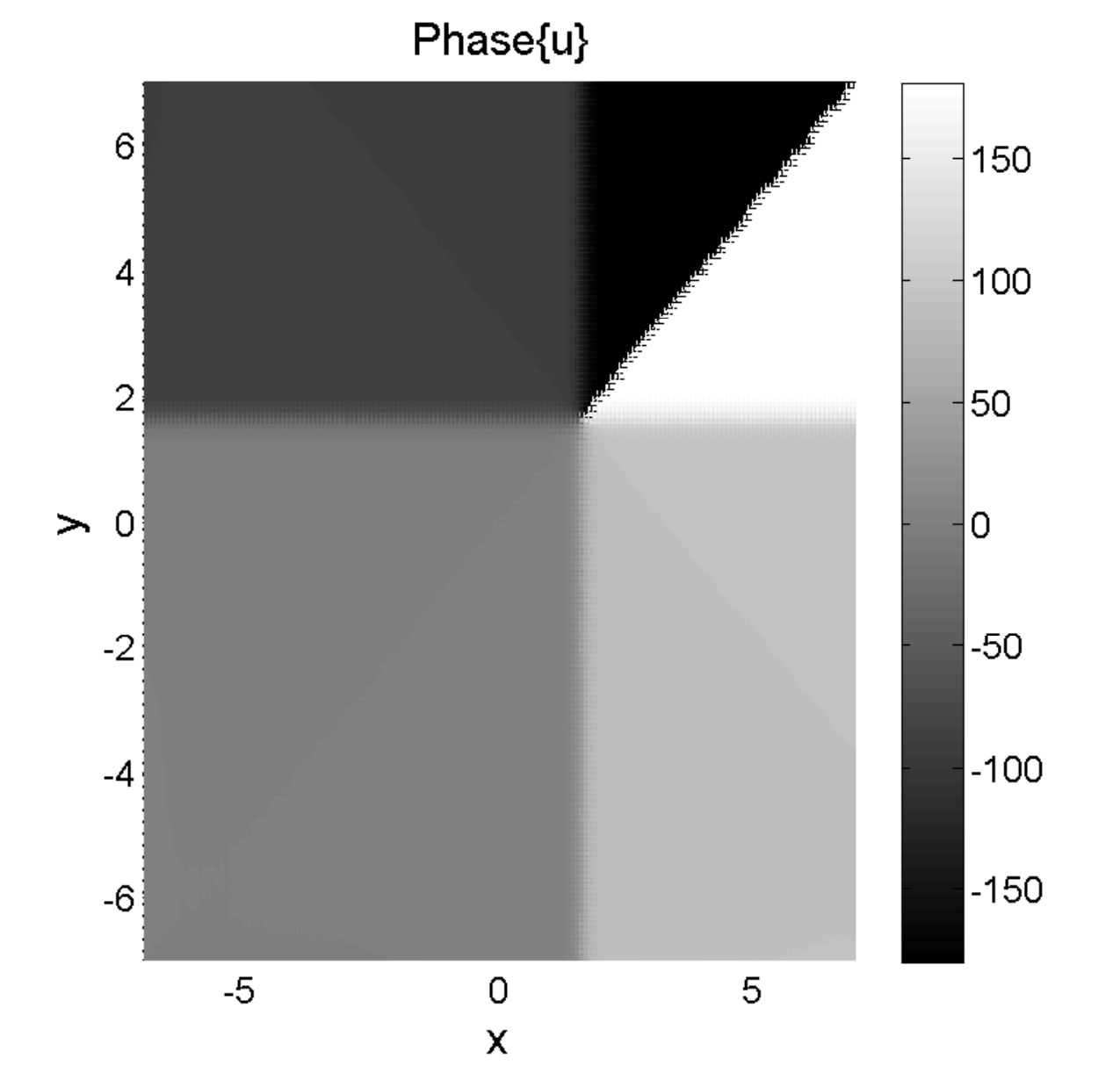}}
\label{SquarePhaseProfile}}
\caption{Generic examples of the amplitude and phase profiles for stable four-peak vortices of the rhombic [panels (a) and (b)] and square-shaped [panels (c) and (d)] types. Both are obtained for $V_{0}=5$ and
$\protect\mu =-7.4$, and belong to the semi-infinite bandgap.}
\label{VortexProfiles}
\end{figure}

In this section we address four-peak vortices of both the rhombic and
square-shaped types (alias the on- and off-sited-centered ones, as said
above), in the semi-infinite bandgap, under the attractive nonlinearity,
assuming that the 2D lattice is subject to the synchronous modulation ($%
\delta =0$). Typical examples of such vortices, in the absence of the OL's time modulation,
where they are known to be stable, are displayed in Fig.~\ref{VortexProfiles}, for $V_{0}=5$ and $\mu =-7.4$ in Eq. (\ref{V05Mum7p4}).

Similar to the analysis presented in Sec. \ref%
{sec:SelfAttractiveSymmetricModulation} for the fundamental solitons,
systematic stability investigation was performed for vortices located
relatively close to the edge of the semi-infinite gap, as well as for ones
found deeper in the bandgap. The respective stability diagrams that
correspond to parameters given in Eq. (\ref{V05Mum7p4}), i.e., in the
vicinity of the bandgap edge, are plotted in panels (a) and (b) Fig.~\ref%
{StabilityDiagram_Vortices}, for the rhombic and square-shaped vortices,
respectively (the respective initial conditions are taken as per Fig.~\ref%
{VortexProfiles}). These two stability diagrams and the one displayed for
the fundamental solitons in Fig.~\ref{StabilityDiagram_1PSin_Wxy0p43N2p087}
demonstrate very close stability patterns.

\begin{figure}[tbp]
\centering
\noindent\makebox[\textwidth]{
\subfigure[]{\includegraphics[width=3.2in]{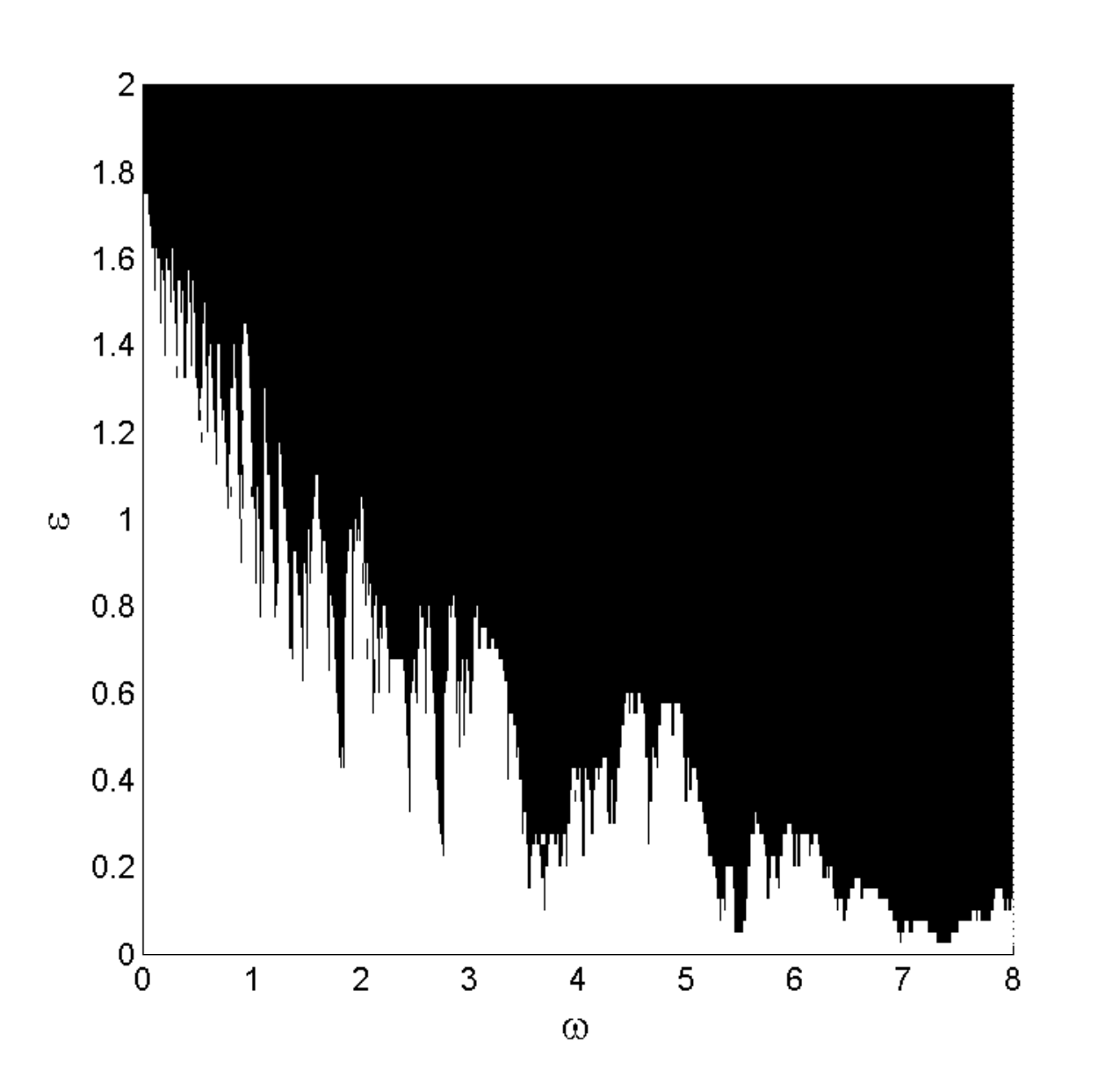}
\label{StabilityDiagram_Rhombic_Mu7p4}}
\subfigure[]{\includegraphics[width=3.2in]{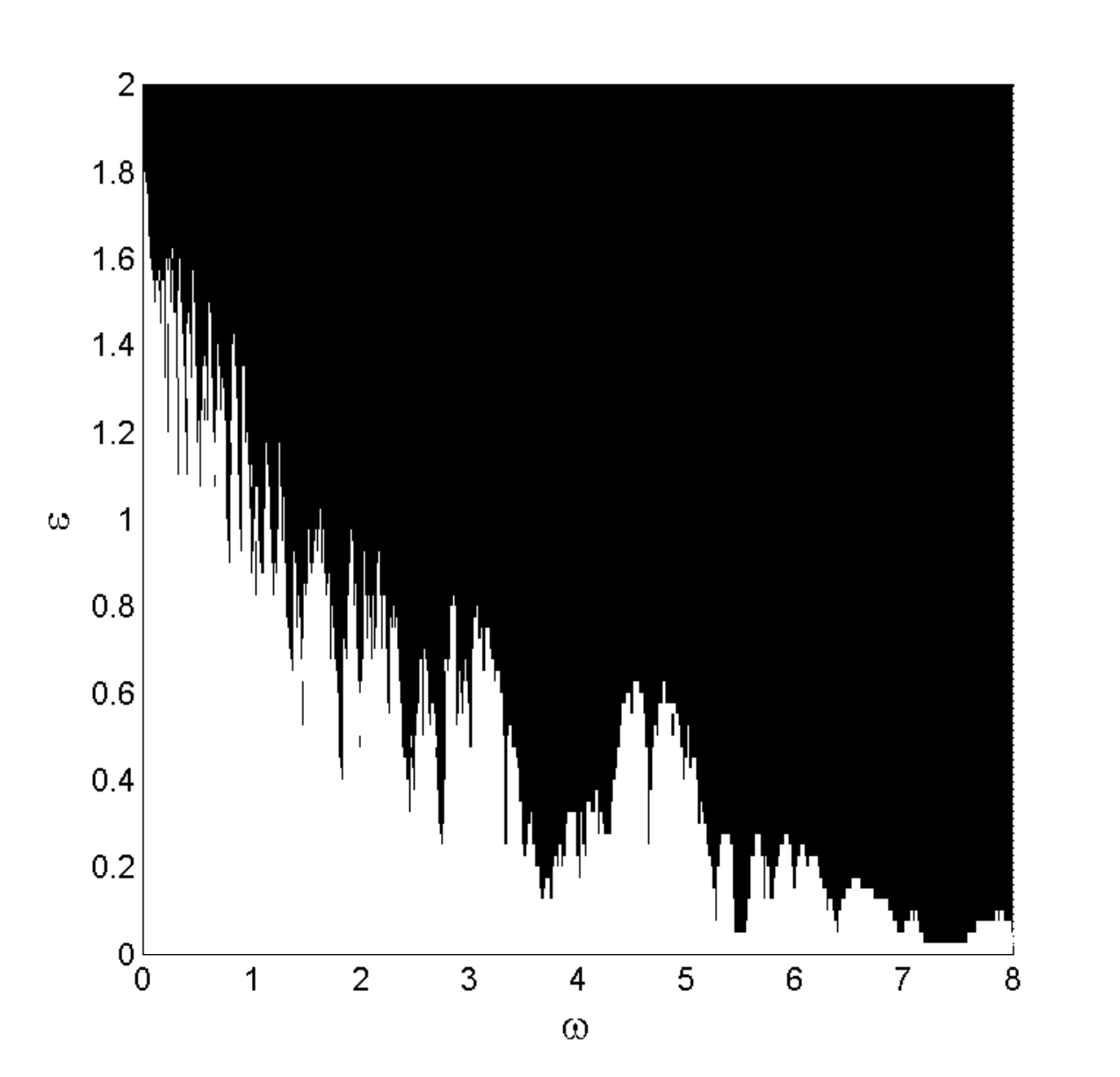}
\label{StabilityDiagram_Square_Mu7p4}}}
\caption{Stability diagrams for 2D vortices of the (a) rhombic and (b)
square-shape types. The result were obtained by means of direct simulations
of Eq. (\protect\ref{2DGPE}), under the self-attractive nonlinearity, with
the synchronous modulation of the 2D lattice ($\protect\delta =0$). In both
cases, the inputs were taken as numerically exact stationary vortex
solutions to the GPE with the static OL potential, for parameters given by
Eq. (\protect\ref{V05Mum7p4}). The respective stationary vortices belong to
the semi-infinite bandgap.}
\label{StabilityDiagram_Vortices}
\end{figure}
Similar conclusions were made for parameters given by Eq. (\ref{V05Mum15}):
the stability diagrams obtained for both types of vortices are very similar
to the one exhibited in Fig.~\ref{StabilityDiagram_1PSin_Wxy0p22N5p29} (not
shown here in detail).

These results demonstrate that the stability patterns of the vortices
practically mimic those for the fundamental solitons, adopting, as the
stability criterion, the $99\%$ norm conservation. This similarity is
explained by the fact that the stability of the four-peak vortex patterns is
chiefly determined by the stability of individual peaks, each being close to
a fundamental soliton, while relatively weak interaction between the peaks
does not introduce additional modes of instability.

\section{Conclusions}

\label{sec:Conclusions} The main objective of this work was to explore the
stability limits of 2D solitons, which are known to be stable under the
action of static 2D OL (optical-lattice) potential, against different
patterns of periodic time-modulation of the lattice. The analysis was
performed by means of systematic simulations of the underlying GPE
(Gross-Pitaevskii equation) model, as well as through the VA (variational
approximation).

First, we have examined the stability of fundamental solitons, which exist
in the semi-infinite bandgap, in the case of the attractive nonlinearity.
Through direct simulations of the GPE we
have found that, when synchronous time modulation is applied to the 2D
lattice as a whole, a structured stability pattern may be identified,
composed of stability peaks which are separated by instability tongues,
which are centered around resonant frequencies. We distinguish two possible
scenarios: the input in the form of the stationary soliton in the static
lattice, chosen close to the edge of the semi-infinite bandgap, where the
stability regions get reduced and distorted, versus inputs placed deep in
the semi-infinite bandgap, that lead to well-structured stability patterns.
Applying the VA method to the synchronous configuration, and appropriately choosing initial parameters for the simulations, we have found that it is possible to produce relatively accurate results. In particular, a good prediction for the resonance frequencies is achieved, with respect to the exact numerical calculations, for initial solitons taken deep in the semi-infinite gap.

The investigation was extended to include different patterns of the
time-modulation, with the phase-shift, $\delta $, between the two 1D
sublattices that form the 2D lattice. We have showed that $\delta =\pi /2$
does not improve the stability and, in fact, makes the stability pattern
somewhat fuzzy. For this scheme, the VA is fairly successful in predicting some of the central resonance frequencies. Applying the shift $\delta =\pi$, we have detected an increase of the stability regions, unless the modulation frequencies are too high. In this case, the VA method could not predict main features of the stability picture and, in particular, large differences were observed when comparing the VA-predicted
and numerically found resonant frequencies.

Applying the synchronous time modulation to fundamental GSs (gap solitons)
in the first bandgap, in the case of the repulsive nonlinearity, the results
demonstrate reduced stability regions, in particular at high modulation
frequencies. In this case, the VA is inaccurate, predicting well-defined
stability patterns, which are not corroborated by the numerical
investigation.

The analysis of the four-peak vortices, of the square-shaped and rhombic
types in the semi-infinite gap, under the action of the attractive
nonlinearity, was also conducted. We have demonstrated that for the
synchronous time modulation applied to 2D lattice as a whole, the resulting
stability diagrams are similar to those constructed for the fundamental
solitons.

\label{sec:Conclusion}

\end{document}